\documentclass{ieeeaccess}
\usepackage{cite}
\usepackage{amsmath,amssymb,amsfonts}
\usepackage{algorithmic}
\usepackage{graphicx}
\usepackage{textcomp}
\usepackage[caption=false,font=footnotesize]{subfig}
\usepackage{enumitem}
\usepackage{here}
\usepackage{url}
\usepackage{footmisc}

\usepackage{bm}
\makeatletter
\AtBeginDocument{\DeclareMathVersion{bold}
\SetSymbolFont{operators}{bold}{T1}{times}{b}{n}
\SetSymbolFont{NewLetters}{bold}{T1}{times}{b}{it}
\SetMathAlphabet{\mathrm}{bold}{T1}{times}{b}{n}
\SetMathAlphabet{\mathit}{bold}{T1}{times}{b}{it}
\SetMathAlphabet{\mathbf}{bold}{T1}{times}{b}{n}
\SetMathAlphabet{\mathtt}{bold}{OT1}{pcr}{b}{n}
\SetSymbolFont{symbols}{bold}{OMS}{cmsy}{b}{n}
\renewcommand\boldmath{\@nomath\boldmath\mathversion{bold}}}
\makeatother

\def\thevol{14}
\def\theyear{2026}

\begin{document}
\history{Received 25 July 2026; accepted 14 August 2026; date of publication 18 August 2026; date of current version 15 August 2026.}
\doi{10.1109/ACCESS.2026.3724865}

\title{Dynamic Tuning of Election Parameters for Timely Leader Failover in State Machine Replication}

\author{\uppercase{Kohya Shiozaki}\authorrefmark{1} and \uppercase{Junya Nakamura}\authorrefmark{2}}

\address[1]{Department of Computer Science and Engineering, Toyohashi University of Technology, Toyohashi, Aichi 441--8580, Japan (e-mail: koya@dsl.cs.tut.ac.jp)}
\address[2]{Information and Media Center, Toyohashi University of Technology, Toyohashi, Aichi 441--8580, Japan (e-mail: junya@imc.tut.ac.jp)}

\tfootnote{This work was supported by JSPS KAKENHI Grant Numbers JP20KK0232 and JP22K11971, and by the Foundation of Public Interest of Tatematsu.}

\markboth
{Shiozaki \headeretal: Dynamic Tuning of Election Parameters for Timely Leader Failover in State Machine Replication}
{Shiozaki \headeretal: Dynamic Tuning of Election Parameters for Timely Leader Failover in State Machine Replication}

\corresp{Corresponding author: Kohya Shiozaki (e-mail: koya@dsl.cs.tut.ac.jp).}

\begin{abstract}
State Machine Replication (SMR) is a technique for achieving fault tolerance in distributed services by replicating the service state across multiple servers.
Leader-based consensus algorithms, e.g., Raft and Multi-Paxos, are commonly used to implement SMR, where one server acts as a leader to coordinate consensus among the others.
When the leader fails, a failover process is triggered to select a new leader.
During this period, called out-of-service (OTS) time, the service temporarily becomes unavailable.
The OTS time depends on the time required for both leader failure detection and election, particularly on election parameters, i.e., the interval of heartbeat messages and the timeout for triggering the election.
If these parameters are improperly configured, the OTS time may increase or a new leader may fail to be elected, resulting in a loss of availability.
Moreover, setting appropriate parameters is difficult under fluctuating network conditions, as suitable parameters change with variations in network latency and packet loss rates.
We propose Dynatune, which dynamically tunes election parameters for timely and stable leader failover according to network conditions by measuring latency and packet loss rates between servers.
Experimental results show that Dynatune reduces leader failure detection time by 78\% and OTS time by 45\% for Raft, and by 79\% and 75\% for Multi-Paxos, while maintaining availability under fluctuating networks.
These results demonstrate that Dynatune effectively enhances failover performance in leader-based SMR.
\end{abstract}


\titlepgskip=-15pt

\maketitle

\section{Introduction}
\label{sec:Introduction}

\PARstart{S}{tate} Machine Replication (SMR)~\cite{schneider_implementing_1990} enhances fault tolerance by replicating the service state across multiple servers.
The servers run a consensus algorithm to agree on a single order of client requests.
These servers then process the requests in that order, so their states remain consistent and the service can continue even if some servers fail.

Many consensus algorithms have been proposed to implement SMR, including Multi-Paxos~\cite{lamport_paxos_1998,lamport_paxos_2001,van_renesse_paxos_2015}, other Paxos-based algorithms~\cite{lamport_fast_2006,zhao_fast_2015,mao_mencius_2008,moraru_there_2013,yan_domino_2020,coelho_geographic_2018,coelho_geopaxos_2021}, and Raft~\cite{ongaro_search_2014}.
Among these, leader-based consensus algorithms such as Multi-Paxos and Raft are widely adopted in practice, where a single server, called the leader, orders client requests while the other servers follow its decisions.

In leader-based consensus algorithms, when the leader fails, the service stops processing new client requests.
To address this, the remaining servers must detect the failure and select a new leader, using mechanisms commonly referred to as heartbeat and leader election.
In this approach, the leader periodically sends heartbeat messages to other servers to indicate that it is still alive.
If the other servers do not receive these messages within a predefined timeout, they suspect that the leader has failed and start a new election.
The duration between the leader's failure and the election of a new leader, during which the service cannot process new requests, is called the \emph{out-of-service (OTS)} time.
Reducing OTS time is therefore crucial for improving the responsiveness and availability of SMR services.

Numerous studies have been conducted to reduce OTS time.
Leaderless or multi-leader consensus algorithms, such as EPaxos~\cite{moraru_there_2013} and Mencius~\cite{mao_mencius_2008}, can achieve high performance under certain workloads by avoiding leader election.
However, their algorithmic and implementation complexity has limited their adoption in production services.
Some works focus on modifying the election mechanism itself to decide a new leader more quickly~\cite{fluri_improving_2018, fu_improved_2021, tang_improved_2022, zhang_escape_2022}.
Other studies investigate election parameters, such as the interval at which the leader sends heartbeat messages and the timeout duration for suspecting leader failure~\cite{iosif_leadership_2020, choumas_using_2022, huang_performance_2020}.
However, these studies mainly examine responsiveness and stability during normal operation, and little is known about how these parameters should be designed or managed to effectively reduce OTS time.

As a result, many services simply adopt the default values\footnote{\url{https://etcd.io/docs/v3.5/tuning/}} or values drawn from practitioner experience.
Such values are not guaranteed to be appropriate, and poor choices can lengthen the OTS time, degrade responsiveness, or even lead to a loss of availability.
Moreover, the appropriate parameters shift with the network conditions and are therefore difficult to keep well configured when the round-trip time~(RTT) or the packet loss rate fluctuates, particularly for SMR deployed over WANs or for replication methods that dynamically move servers closer to client locations~\cite{kostler_fluidity_2023, chiba_state_transfer_2022}.

In this study, we first analyze how election parameters should be configured to reduce OTS time and then propose Dynatune, a mechanism that dynamically tunes these parameters.
Dynatune extends the heartbeat messages already used by leader-based consensus algorithms for failure detection, enabling each server to estimate RTT and packet loss rates in real time and apply election parameters derived from these estimates.
This allows the parameters to remain appropriate under fluctuating network conditions without manual tuning.
Since Dynatune only extends heartbeat messages and leaves the core consensus mechanism unchanged, it preserves the correctness of the original algorithm.

We implement Dynatune on etcd\footnote{\label{footnote:etcd_url}\url{https://github.com/etcd-io/etcd}}, a widely used key-value store based on Raft.
To verify its applicability beyond Raft, we port it to JPaxos~\cite{konczak_jpaxos_2011}, an open-source implementation of Multi-Paxos.
Both implementations are publicly available: the Raft version\footnote{\label{footnote:raft_url}\url{https://github.com/distsys-lab/dynatune}} and the Multi-Paxos version\footnote{\label{footnote:dynatune_url}\url{https://github.com/distsys-lab/JPaxos}}.
Dynatune reduces the failure detection time and OTS time by 78\% and 45\% on Raft, and by 79\% and 75\% on Multi-Paxos, while maintaining availability under fluctuating network conditions.

Our key contributions are as follows.
\begin{enumerate}[leftmargin=*]
\item We analyze how network conditions affect election parameters in leader-based SMR and clarify the conditions required to reduce OTS time without sacrificing availability.
\item We propose Dynatune, which extends heartbeat messages to measure latency and packet loss rates between servers and dynamically tunes election parameters based on these measurements, enabling timely leader failover even under fluctuating network conditions.
\item We provide open-source implementations of Dynatune for two representative leader-based consensus algorithms: one for Raft, built on etcd, and one for Multi-Paxos, built on JPaxos.
\end{enumerate}

The remainder of this paper is organized as follows.
Section~\ref{sec:Background} introduces leader failover in SMR and the impact of network conditions on election parameters.
Section~\ref{sec:Dynatune} introduces Dynatune and describes its design.
Section~\ref{sec:Evaluation} comprehensively evaluates Dynatune using the Raft-based implementation.
Section~\ref{sec:Applicability} then demonstrates that Dynatune is also applicable to other leader-based consensus algorithms by applying it to Multi-Paxos.
Section~\ref{sec:Related_Work} reviews the related work.
Finally, Section~\ref{sec:Conclusion} concludes the study.

A preliminary version of this work has been reported in~\cite{shiozaki_dynatune_2025}.
This manuscript extends the conference version by applying Dynatune to Multi-Paxos, implementing it on JPaxos and evaluating it experimentally (Section~\ref{sec:Applicability}).
This demonstrates that Dynatune is applicable to leader-based consensus algorithms beyond Raft.
It also revises Section~\ref{sec:Related_Work} to discuss BALLAST~\cite{wang_ballast_2025}, which is a preprint released after the publication of the conference version and shares our motivation of reducing OTS time by adaptively controlling election timeout.
In addition, all experiments that did not use Amazon Web Services (AWS) were rerun with a corrected hardware configuration, and experimental data and discussion omitted from the conference version due to page constraints were supplemented.

\section{Background}
\label{sec:Background}
\subsection{State Machine Replication}

SMR is a technique that replicates the service state to enhance fault tolerance.
In this study, we focus on \emph{non-Byzantine} failures, where servers may crash and later recover but do not exhibit malicious behavior.
Under this assumption, SMR replicates the service state across $2f+1$ servers, allowing the service to continue operation with a consistent state even if up to $f$ servers fail.
This is achieved by using a consensus algorithm that ensures a majority of servers agree on the order of requests.

Many SMR services adopt leader-based consensus algorithms, including Raft and Multi-Paxos, because of their simplicity and clear structure.
In this study, we use Raft as a representative example to illustrate our discussion, implementation, and evaluation of the proposed method.
This is because Raft explicitly specifies the leader failover process as part of its algorithm, and its correctness has been formally verified~\cite{ongaro_search_2014}.
Note that other leader-based consensus algorithms, such as Multi-Paxos, share mostly the same failover mechanism as Raft; thus, the proposed method is also applicable to these algorithms.
We discuss this applicability in detail in Section~\ref{sec:Applicability}.

A Raft server is always in one of three roles, namely \emph{leader}, \emph{follower}, or \emph{candidate}.
During normal operation, one server acts as the leader, and the others act as followers.
A client request first reaches the leader, which appends it to its log and forwards the entry to the followers.
Each follower stores the entry and returns an acknowledgment.
After a majority of the servers have persisted the entry, the leader commits it and answers the client.
Every server subsequently applies committed entries to its local state machine in log order, so the service state stays consistent across servers and the service keeps running even when some followers fail.

\subsection{Leader Failover in Raft}

\begin{figure*}[tb]
    \centering
    \includegraphics[width=0.8\textwidth]{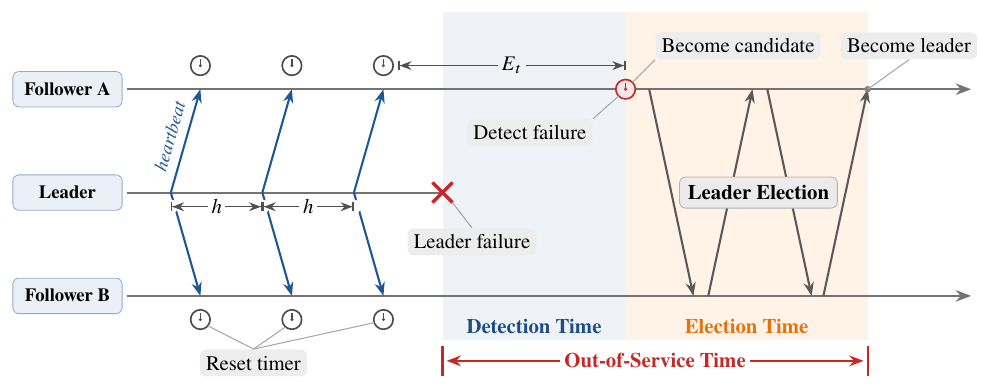}
    \caption{Leader failover process in Raft.}
    \label{fig:raft_ElectionIllustration}
\end{figure*}

Raft stops processing new requests when its leader fails.
To address this, Raft uses heartbeats and leader elections to select a new leader, as shown in Figure~\ref{fig:raft_ElectionIllustration}.
The leader repeatedly sends heartbeats to all other servers at an interval called the \emph{Heartbeat Interval}~$ h $.
Followers have an election timer of length \emph{Election Timeout} $ E_t $ and reset it when receiving a heartbeat from the leader.
When the leader fails and its heartbeats stop, the election timer of a follower expires, and the follower determines that the leader has failed.
The follower then increments Raft's \emph{term}, a logical election counter, becomes a candidate, and immediately starts an election for a new leader.
If the candidate receives votes from the majority of the servers, it becomes the new leader and resumes processing requests.
We refer to the duration from the leader failure until a follower detects the failure as the \emph{detection time}, and to the duration from that detection until a new leader is elected as the \emph{election time}.
The OTS time is the sum of these two durations.

If several followers become candidates around the same time, the votes are divided among them, and none obtains a majority because each server votes only once per election.
In this situation, known as a \emph{split vote}, candidates must wait for their election timers to expire again before starting a new election.
This repetition of elections continues until a leader is elected, and increases the OTS time.
To decrease the frequency of split votes, each server randomizes its election timeout based on $E_t$ whenever its timer expires.
We refer to the actual timer value after randomization as $\mathrm{randomizedTimeout}$.

\subsection{Election Parameters and Network Conditions}
\label{sec:election_parameters_and_network_conditions}
The election parameters are critical to the availability and responsiveness of the service.
However, the methodology for appropriately configuring these parameters is not well understood.
According to the original Raft study~\cite{ongaro_search_2014}, election timeout $E_t$ should be set to satisfy the following inequality:
\begin{equation*}
\mathrm{broadcastTime} \ll E_t \ll \mathrm{MTBF}
\end{equation*}
Here, $\mathrm{broadcastTime}$ denotes the average time for a server to send Remote Procedure Calls (RPCs) to all other servers in parallel and collect their responses.
In WAN environments, this value is treated as equal to the maximum RTT between servers because network delay typically dominates per-server processing overheads.
$\mathrm{MTBF}$ refers to the mean time between failures of a single server.
The heartbeat interval $h$ is set to be significantly smaller than $E_t$.
Heartbeats then keep arriving within the timeout, which prevents followers from starting unnecessary elections and thus preserves availability.

Inappropriate election parameter settings may lead to significant issues.
The RTT between servers, particularly between the leader and the followers, affects $ E_t $.
If $E_t$ is smaller than the RTT, a follower's timer expires before the next heartbeat arrives.
Followers then repeatedly start elections while requests remain unprocessed, and availability is lost.
Conversely, if $E_t$ is too long, the leader's failure detection is delayed, increasing the OTS time.

For the heartbeat interval, Huang et al.~\cite{huang_performance_2020} recommended shortening $h$ so that more heartbeats, $K$~$(= E_t/h)$, are sent before the timer expires.
However, frequent heartbeats strain the leader's computational resources and network bandwidth, thus reducing request-processing capability.

Therefore, the election parameters must be carefully configured according to network conditions, such as the RTT and packet loss rate.

\subsection{Network Fluctuations and Problems in WAN Environments}
\label{sec:network_fluctuations_problem}
\begin{figure}[tb]
    \centering
    \includegraphics[width=0.8\columnwidth]{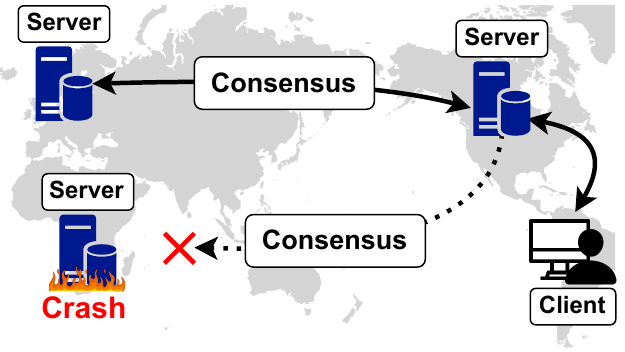}
    \caption{Geo-replicated SMR with three servers.}
    \label{fig:geo-smr}
\end{figure}

Modern SMR services are not limited to LAN environments and are often deployed across multiple geographically distant data centers.
Such a configuration is termed geo-replicated SMR~\cite{coelho_geographic_2018, yan_domino_2020, moraru_there_2013, mao_mencius_2008, kostler_fluidity_2023,shiozaki_selection_2025}, as depicted in Figure~\ref{fig:geo-smr}.
Geographic dispersion allows the service to maintain availability during large-scale disruptions such as earthquakes, tsunamis, or region-wide power failures.
Geo-replicated SMR can also improve performance for globally distributed users by placing servers closer to them.

The cost of this dispersion is that the RTT and packet loss rate tend to increase with inter-server distance and, more importantly, do not stay constant.
These fluctuations can arise from ordinary usage patterns, including daily traffic cycles and large-scale events, or from failures and operational events, such as submarine cable failures, scheduled maintenance, and public-cloud outages that change routing paths.

\subsubsection{RTT Fluctuation}
Several studies report that WAN RTT varies widely over time as routes and congestion levels change~\cite{rtt_2017_alves, latency_2016_hj, rtt_2024_martinez}.
Høiland-Jørgensen et al.~\cite{latency_2016_hj} observed RTT variation exceeding 100\,ms on more than two-thirds of the paths they sampled and reported that increased load can raise queuing delays beyond 200\,ms, indicating that the variation is substantial.

Such variability cannot be accommodated by a statically chosen $E_t$.
Suppose that $E_t$ is set to the maximum RTT observed under normal conditions so that failures are detected quickly.
If the RTT later exceeds this value, followers repeatedly start elections even though the leader has not failed, and the service remains unavailable until the RTT falls back below $E_t$.
Conversely, enlarging $E_t$ to tolerate such increases delays the detection of genuine failures and increases the OTS time.
Static settings cannot resolve this trade-off, which motivates dynamic adjustment.

\subsubsection{Packet Loss Rate Fluctuation}
Packet loss rates in WANs fluctuate as well.
Haq et al.~\cite{haq_measuring_2017} reported that dedicated inter-data-center connections keep packet loss rates low, between 0.001\% and 0.1\%, whereas paths over the public Internet range from 0.001\% up to 50\%.
Mok et al.~\cite{mok_measuring_2021} further reported that packet loss rates on inter-cloud paths within Google Cloud Platform can rise from 3\% to 50\% during regional congestion.

The heartbeat interval $h$ is normally configured for an assumed packet loss rate so that at least $K$ heartbeats are delivered within one timeout window.
When the actual packet loss rate exceeds this assumption, too few heartbeats reach the followers in time, causing their timers to expire and unnecessary elections to occur.
Setting $h$ very small in advance can tolerate spikes in the packet loss rate, but it increases the leader's CPU load and network-bandwidth consumption, reducing its request-processing capacity.
As with RTT, a fixed $h$ cannot resolve this trade-off under fluctuating packet loss rates.

\section{Dynatune}
\label{sec:Dynatune}

This section presents Dynatune, which measures the network in real time and dynamically tunes election parameters to reduce the OTS time safely and effectively.
Section~\ref{sec:assumptions} first states the assumptions, and Section~\ref{sec:overview} gives an overview of Dynatune.
Section~\ref{sec:network_measurement} describes how the network is measured through heartbeats, Section~\ref{sec:optimization} explains how the election parameters are derived from the measurements, and Section~\ref{sec:implementation} presents the implementation on Raft.

\subsection{Assumptions}
\label{sec:assumptions}

Dynatune operates under the following assumptions equivalent to those of Raft~\cite{ongaro_search_2014} because it does not modify Raft's fundamental mechanisms, particularly the consensus on the request execution order and leader election.

We assume an asynchronous network model with non-Byzantine failures.
Servers may crash and later restart, and messages may be delayed, lost, duplicated, or reordered, but they are not corrupted.
The system does not assume clock synchronization or bounded message delays.

Under these assumptions, we consider a cluster consisting of $2f+1$ servers.
As long as the number of failed servers does not exceed $f$, the system provides the following guarantees.

\begin{itemize}
  \item \emph{Safety} is guaranteed. Once a value is decided, no two non-faulty servers ever decide different values, and the decision is never revoked.

  \item \emph{Liveness} is not guaranteed during purely asynchronous periods. When the system eventually experiences sufficiently long periods during which the non-faulty servers can communicate with bounded (but unknown) message delays, liveness is guaranteed; that is, liveness is guaranteed under partial synchrony.
\end{itemize}

The assumptions and guarantees described above are as strong as those of Raft and Multi-Paxos.
In Multi-Paxos, liveness is fully guaranteed by introducing additional weak assumptions, such as finite bounds on clock drift and message delays.
Under the same class of weak assumptions, liveness is also guaranteed in Raft and Dynatune.
Therefore, the strength of the assumptions required to guarantee safety and liveness is the same for Raft, Multi-Paxos, and Dynatune.

\subsection{Overview}
\label{sec:overview}

The core idea of Dynatune is to lower the election timeout $E_t$ as far as safety allows so that leader failures are detected quickly.
Dynatune keeps the election parameters well configured through real-time network measurements, even while the network fluctuates.

Dynatune adjusts the election parameters through the following cycle:
\begin{enumerate}[label=\textbf{Step \arabic*:}, start=0, leftmargin=*, labelindent=\parindent]
    \item The leader repeatedly sends heartbeats with additional metadata for measuring network metrics to followers. Each follower records metadata when it receives the heartbeats. If the follower has received sufficient metadata, it moves to Step 1. Note that after moving to Step 1, each follower continues to record metadata.
    
    \item Each follower estimates the RTT and packet loss rate between itself and the leader using the recorded metadata.
    
    \item Each follower determines $E_t$ based on the RTT and then calculates $h$ from the packet loss rate.
    
    \item Each follower applies the updated $E_t$ to its election timer and sends the leader the updated $h$ piggybacked in the heartbeat response.
    The leader applies the received $h$ as the heartbeat sending interval to the corresponding follower.
\end{enumerate}
Dynatune follows changing network conditions by repeating Steps 1--3.
Whenever a follower starts an election process, whether due to a leader failure or a sudden rise in the RTT, it discards its collected RTT and packet loss data, and the cycle restarts from Step 0.
Step 0 begins with the election parameters at their default values.
These defaults are large and conservative.
Thus, even if tuning fails and temporarily results in $E_t < \mathrm{RTT}$, falling back to the defaults maintains availability.

Raft and Multi-Paxos ordinarily use one identical set of election parameters for every server.
Dynatune instead tunes the parameters separately for each leader-follower communication path.
Each path is tuned according to its own network conditions, which maximizes the effect of tuning.
The measurement and tuning for each path are performed by the followers rather than the leader, because the leader is more heavily loaded than the followers.
This design mitigates the impact of tuning overhead and preserves request-processing capacity.

Dynatune embeds the information needed for network measurement and parameter tuning in heartbeat messages.
This design avoids additional communication among servers and keeps the implementation simple.
The details of the embedded fields are described in Section~\ref{sec:network_measurement}.

\subsection{Network Metrics Measurement}
\label{sec:network_measurement}
This section explains how Dynatune measures the RTT and packet loss rate through heartbeats.

\begin{figure*}[tb]
    \centering
    \includegraphics[scale=0.76]{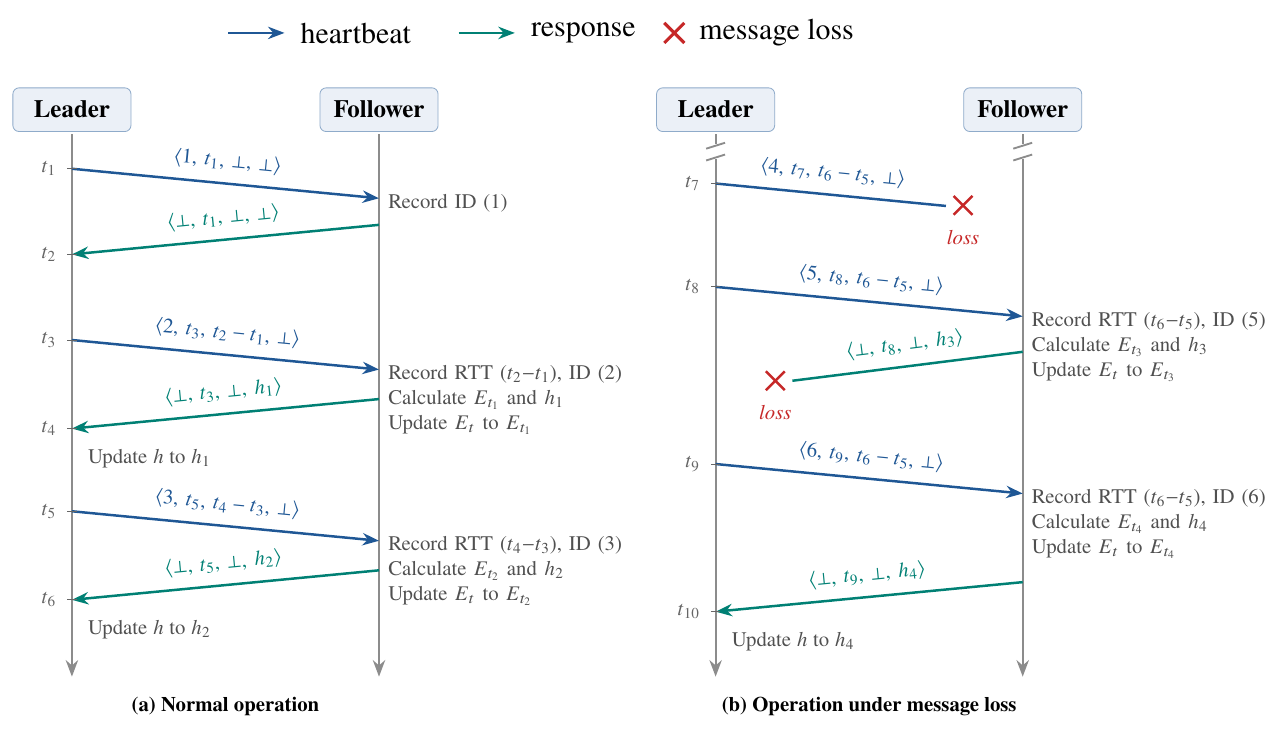}
    \caption{Overview of Dynatune's extended heartbeat mechanism for network measurement and election parameter adaptation. Both leader-to-follower and follower-to-leader messages use the same format $\langle \mathrm{ID}, \mathrm{ts}, \mathrm{RTT}, h \rangle$. Fields marked with $\bot$ denote null values.}
    \label{fig:network_measurement}
\end{figure*}

Dynatune extends the default Raft heartbeat message with additional fields.
Figure~\ref{fig:network_measurement} provides an overview of the extended heartbeat mechanism.
The extended heartbeat message is conceptually represented as $\langle \mathrm{ID}, \mathrm{ts}, \mathrm{RTT}, h \rangle$.
Here, $\mathrm{ID}$ denotes a monotonically increasing heartbeat identifier assigned by the leader,
and $\mathrm{ts}$ denotes the timestamp recorded by the leader when sending the heartbeat.
$\mathrm{RTT}$ denotes the round-trip time measured from the previous heartbeat exchange and delivered to the follower.
$h$ denotes a heartbeat interval value computed by the follower and fed back to the leader.

Each follower maintains two lists: $\mathrm{RTTs}$ for storing RTT values and $\mathrm{ids}$ for storing heartbeat IDs.
Upon receiving each heartbeat, the follower adds the RTT to $\mathrm{RTTs}$ and the corresponding ID to $\mathrm{ids}$, thereby collecting the data necessary to measure the RTT and packet loss rates.
The specific measurement methods are described below.

\subsubsection{RTT}
\label{sec:rtt_measurement}

The leader measures the RTT by attaching a timestamp to each outgoing heartbeat and measuring the elapsed time when the follower's response arrives.
Concretely, the leader writes the sending timestamp into the heartbeat and sends it.
The follower replies immediately upon receipt, and the leader obtains the RTT as the difference between the send and receive timestamps.
The leader then delivers this RTT to the follower on the next heartbeat, and the follower stores it in $\mathrm{RTTs}$ and updates $E_t$ by the method described in Section~\ref{sec:et_optimization}.

This measurement method relies only on the leader's local clock, enabling accurate RTT measurements without assuming clock synchronization across servers.
In this design, all information required for RTT computation is embedded in the heartbeat messages, so the leader does not retain any state for computation.
In other words, the leader does not need to track measurement state or use timeouts to decide when to discard it, even if a heartbeat is lost.
Furthermore, RTTs can be computed correctly even when the heartbeat sending order does not correspond to its arrival order.

\subsubsection{Packet Loss Rate}

The leader assigns a sequential ID to each heartbeat to measure the packet loss rate.
Each follower has a list $\mathrm{ids}$ of the received heartbeat IDs.
The follower then calculates the packet loss rate $p$ with $\mathrm{ids}$ as follows:
\begin{equation*}
p = 1 - \frac{\operatorname{length}(\mathrm{ids})}{\mathrm{ids}[-1] - \mathrm{ids}[0] + 1}
\end{equation*}

As described in Section~\ref{sec:assumptions}, the arrival order of heartbeat messages is not guaranteed, and packets may be duplicated.
To address this, the follower inserts the IDs into the list in ascending order and ignores subsequent receptions when duplicate messages with the same ID arrive.

\subsection{Tuning of Election Parameters}
\label{sec:optimization}

This section explains how Dynatune tunes the election parameters using the RTT and packet loss rate measured in Section~\ref{sec:network_measurement}.
The parameters are the follower's election timeout $E_t$ and the leader's heartbeat interval $h$.
Dynatune first determines $E_t$ from the RTT, and then adjusts $h$ according to the packet loss rate $p$.
As discussed in Sections~\ref{sec:election_parameters_and_network_conditions} and~\ref{sec:network_fluctuations_problem}, choosing appropriate values of $E_t$ and $h$ is important for the availability and responsiveness of an SMR service.

\subsubsection{Election Timeout}
\label{sec:et_optimization}

$E_t$ must be larger than the RTT, as described in Section~\ref{sec:election_parameters_and_network_conditions}, while remaining as small as possible to shorten the detection time.
Ideally, setting $E_t = \mathrm{RTT}$ would minimize the detection time.
However, RTT fluctuations must also be taken into account.
Thus, Dynatune sets $E_t$ based on the statistical characteristics of the RTT between the leader and follower to balance shorter detection time against the risk of false detection.
Specifically, we calculate $E_t = \mu_{\mathrm{RTT}} + s\sigma_{\mathrm{RTT}}$,
where $\mu_{\mathrm{RTT}}$ is the average RTT between the leader and follower, and $\sigma_{\mathrm{RTT}}$ is its standard deviation.
These values are calculated using the $\mathrm{RTTs}$ list described in Section~\ref{sec:rtt_measurement}.
The coefficient $s$ is a safety factor set in advance by the practitioner.
A larger $s$ makes $E_t$ tolerate larger RTT fluctuations and reduces the risk of false detections caused by network delay fluctuations.
When a new RTT is measured, $E_t$ is recalculated using the updated average RTT and standard deviation.
This approach reduces the risk of false detections while shortening the detection time.

\subsubsection{Heartbeat Interval}
\label{sec:h_optimization}

We determine $h$ from $E_t$ and the packet loss rate between the leader and follower, considering the trade-off between resource consumption and the risk of false detection.
First, we calculate the number of heartbeats $K$ needed so that the probability that a follower receives at least one heartbeat is greater than or equal to $x$, taking the packet loss rate into account.
We then choose $h$ so that $K$ heartbeats are sent at equal intervals within $E_t$.

Specifically, for packet loss rate $p$, the condition that $K$ must satisfy to meet the arrival probability $x$ is $1 - p^K \geq x$.
From this condition, we obtain the required number of heartbeats $K$ as follows:
\begin{equation*}
K =
\begin{cases}
1 & \text{if $p = 0$,} \\
\left\lceil \log_p(1-x) \right\rceil & \text{if $0 < p < 1$.}
\end{cases}
\end{equation*}
Finally, to send $K$ heartbeats within a period $E_t$, we set $h = E_t/K$.
For consensus algorithms that randomize the election timeout, such as Raft, the follower uses its actual timer value, $\mathrm{randomizedTimeout}$, in place of $E_t$ in this calculation.
This $h$ is recalculated each time a heartbeat is received, and the updated value of $h$ is piggybacked on the response to the leader, which then applies it.
This method ensures that the follower receives at least one heartbeat with high probability before its election timer expires, thereby reducing both the risk of unnecessary elections owing to false detections and resource consumption.

\subsection{Implementation on Raft}
\label{sec:implementation}

We implemented Dynatune on etcd\footref{footnote:etcd_url}, a widely used open-source key-value store that employs the Raft consensus algorithm.
We forked etcd and integrated Dynatune into it; the source code is publicly available online\footref{footnote:dynatune_url}.

Many production Raft implementations, including etcd, provide a pre-vote phase before starting an election to improve stability.
In this phase, a candidate first conducts a preliminary vote and starts the actual election only if a majority of the servers approve it.
A server approves the pre-vote only if its own election timer has expired, indicating that it suspects a leader failure.
This prevents a follower that is temporarily disconnected from a live leader from causing unnecessary term increments and elections, thereby avoiding needless leader changes.
Our Raft-based implementation of Dynatune retains this phase within the election process that etcd provides.

The original etcd implementation uses TCP for all Raft communication.
Dynatune sends heartbeat messages over UDP to reduce their communication cost, while retaining TCP for the other consensus-related communication.

In addition, we added runtime arguments to configure Dynatune's parameters.
These arguments specify the safety factor $s$ and the arrival probability $x$ used to calculate $E_t$ and $h$, respectively.
Followers can also configure the list size for recording network metrics, specifying both minimum ($\mathrm{minListSize}$) and maximum ($\mathrm{maxListSize}$) sizes.
If the list size is below $\mathrm{minListSize}$, Dynatune remains in \textbf{Step~0}, and the followers continue recording the metadata until sufficient data are stored.
Once the list size reaches $\mathrm{minListSize}$, Dynatune transitions to \textbf{Step~1} and begins tuning the election parameters.
When the list size exceeds $\mathrm{maxListSize}$, Dynatune discards the oldest data to prevent the list from growing indefinitely.

\section{Evaluation}
\label{sec:Evaluation}
This section evaluates the Raft-based implementation of Dynatune under various fault scenarios and network conditions.
Section~\ref{sec:exp_setting} first describes the common experimental settings.
Sections~\ref{sec:experiments_under_stable_network} and~\ref{sec:experiments_under_fluctuating_network} then present the experiments under stable and fluctuating networks, respectively, followed by the real distributed experiment conducted on AWS in Section~\ref{sec:distributed_experiment} and the discussion in Section~\ref{sec:discussion}.

\subsection{Experimental Settings}
\label{sec:exp_setting}

We conducted the experiments on a single physical machine, reproducing a geographically distributed environment for all experiments except the one described in Section~\ref{sec:distributed_experiment}.
Specifically, we deployed multiple Docker containers, treated each container as one server of a key-value store cluster, and injected delays and packet losses into each container's traffic with the \texttt{tc} command.
We used the etcd implementation for the key-value store, referring to the default etcd as \emph{Raft} and to the Dynatune-integrated version described in Section~\ref{sec:implementation} as \emph{Dynatune}.

In these experiments, we extracted from each server's log files the time of the leader's failure, the time when the failure was detected, and the time when a new leader was elected in order to calculate the detection and OTS times.
Running the servers on multiple physical machines would make these measurements inaccurate because hardware clocks are not perfectly synchronized across machines.
We therefore ran all servers on a single machine so that they share one hardware clock.
This made it possible to calculate the detection and OTS times accurately from the log timestamps of different servers.

The machine primarily used for the experiments was equipped with an Intel Xeon Silver 4310 2.10\,GHz CPU with 12 cores, with each Docker container allocated four cores and 4\,GB of memory.
We used a 1\,TB NVMe SSD for storage without imposing any I/O limits on the containers.

We configured the election parameters for Raft with default values of $E_t = 1000$\,ms and $h = 100$\,ms.
Dynatune uses the same default election parameters as Raft.
While these parameters are changed dynamically during normal operation, Dynatune falls back to the default parameters when a leader election is triggered.
We also configured the four runtime arguments of Dynatune as follows: $s = 2$, the probability $x = 0.999$, $\mathrm{minListSize} = 10$, and $\mathrm{maxListSize} = 100$\footnote{The conference version incorrectly stated $\mathrm{maxListSize} = 1000$; the actual value used in the experiments was 100.}.

\subsection{Experiments under Stable Network Conditions}
\label{sec:experiments_under_stable_network}
Under stable network conditions, we evaluated failover performance and runtime overhead.
Specifically, we fixed the RTT between the servers at 100\,ms and the packet loss rate $p$ at 0\% without intentionally introducing jitter.

\subsubsection{Failover Performance}
\label{sec:election_performance}
First, we validated the effectiveness of the proposed method by intentionally failing the leader 1000 times across five servers and measuring the detection and OTS times for Raft and Dynatune to evaluate their failover performance.
A leader failure was emulated by suspending the leader's container.

\begin{figure}
    \centering
    \includegraphics[width=0.9\linewidth]{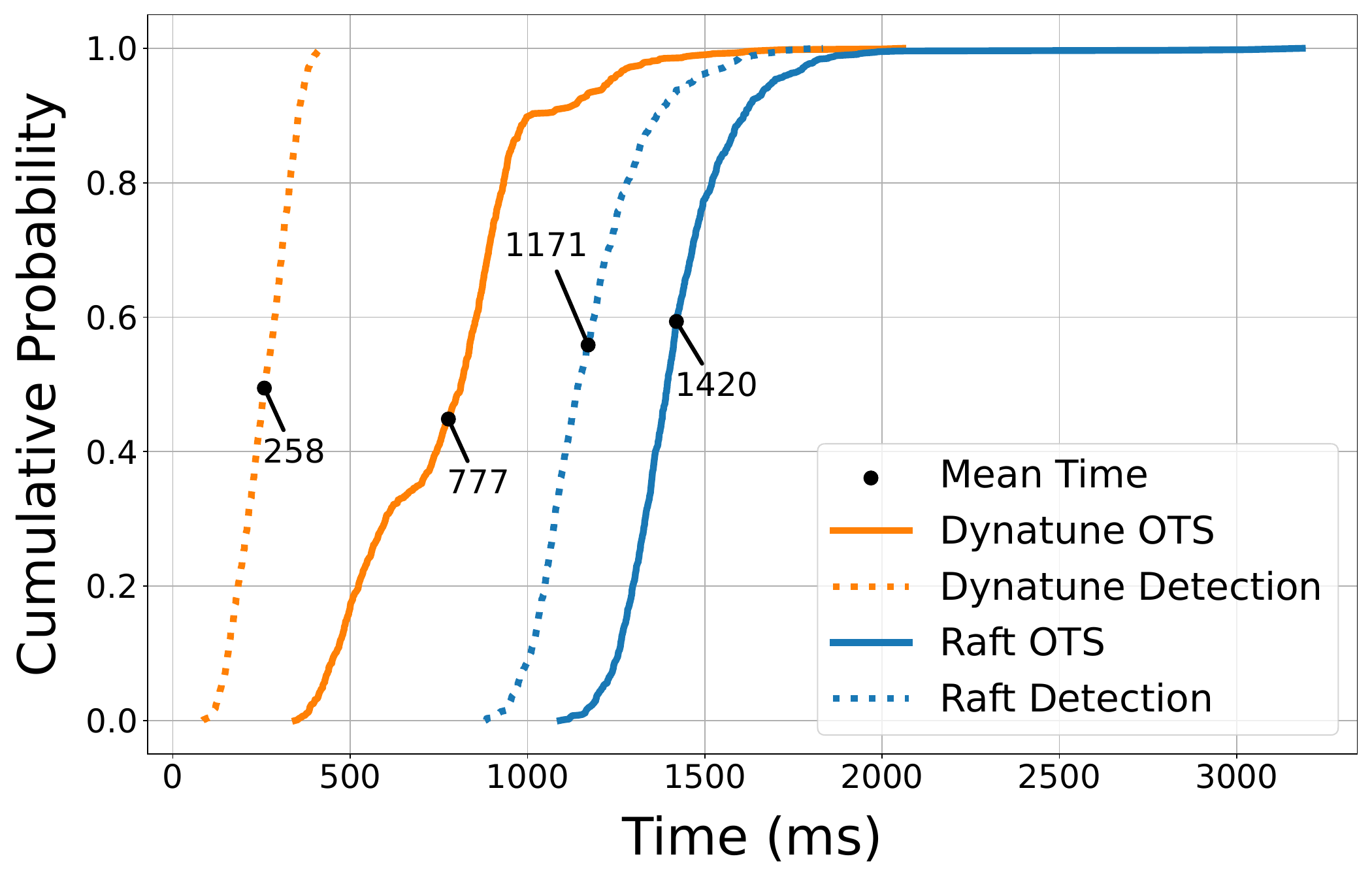}
    \caption{Cumulative probability of detection and OTS times for Dynatune vs.~Raft.}
    \label{fig:election_performance_exp}
\end{figure}

Figure~\ref{fig:election_performance_exp} shows the results.
Focusing on the mean values, Dynatune reduced the detection time by 78\%, from 1171\,ms to 258\,ms, compared with Raft.
Similarly, Dynatune reduced the OTS time by 45\%, from 1420\,ms to 777\,ms.
Additionally, the mean $\mathrm{randomizedTimeout}$ at the moment of detection was 1454\,ms for Raft but only 203\,ms for Dynatune.
This indicates that the smaller $E_t$ of Dynatune enabled the faster detection and shorter OTS time.

\begin{figure}
    \centering
    \includegraphics[width=0.8\linewidth]{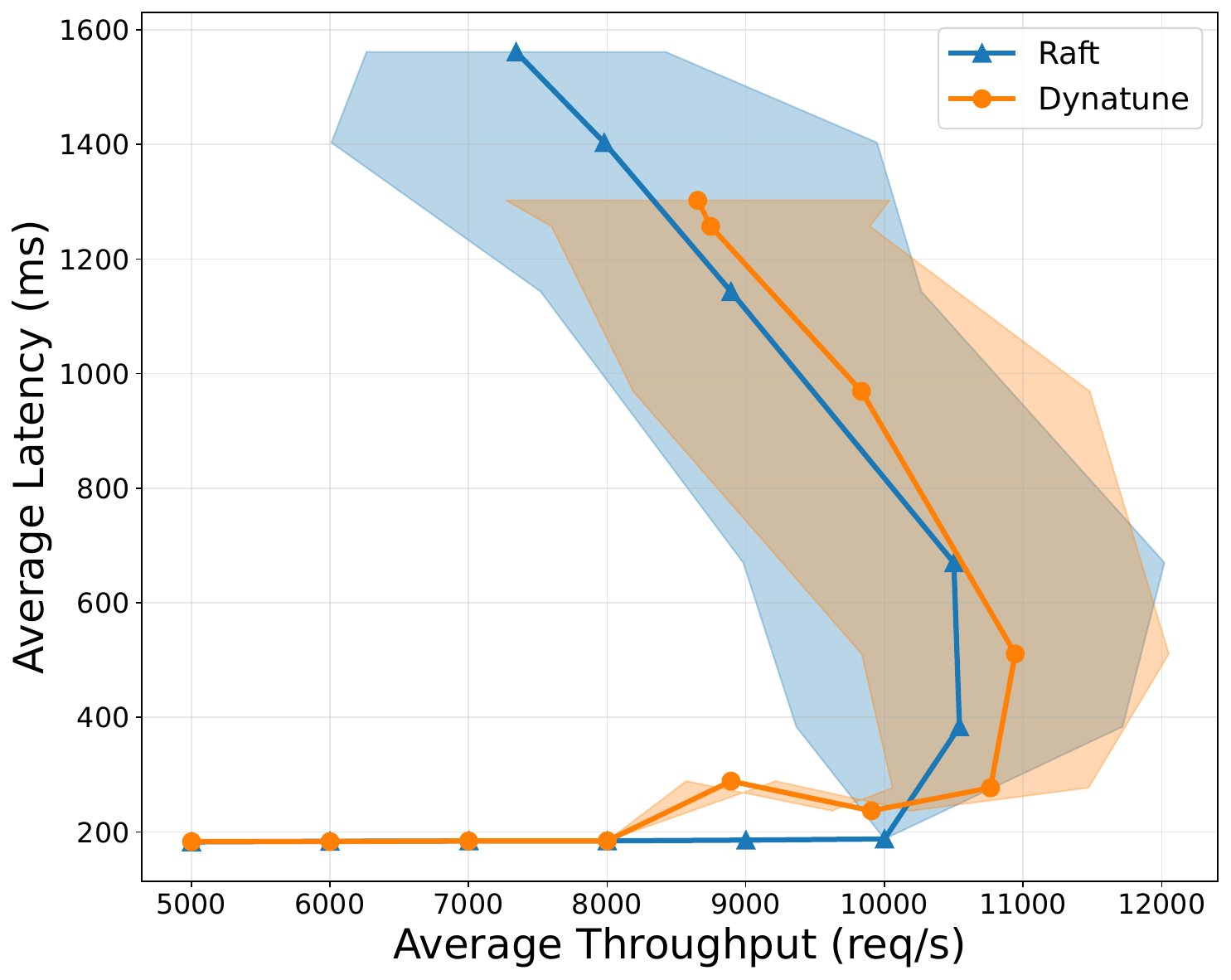}
    \caption{Throughput and latency for Dynatune vs.~Raft.}
    \label{fig:peak_throughput}
\end{figure}

\subsubsection{Peak Throughput without Failures}
\label{sec:peak_throughput}

Next, we measured the peak throughput to quantify the overhead of Dynatune's automatic tuning.
The experimental setup was identical to that described in Section~\ref{sec:election_performance}; however, no failures were induced.
Instead, we introduced clients and increased the load stepwise until the throughput peaked.
The clients issued requests in an open loop.
We increased the offered load in steps of 1000\,req/s, held each step for 10\,s, and repeated the entire measurement ten times.

Figure~\ref{fig:peak_throughput} plots the average latency, the average throughput, and the throughput standard deviation.
At their peaks, Raft reached 10541\,req/s and Dynatune reached 10945\,req/s, indicating that their peak throughputs are comparable\footnote{This result differs in trend from the conference version~\cite{shiozaki_dynatune_2025}, in which the peak throughput of Dynatune was 6.4\% lower than that of Raft. We attribute this gap to a hardware fault that destabilized the original experimental environment. After repairing the hardware and re-running the measurements, the gap disappeared.}.

\subsection{Experiments under Fluctuating Network Conditions}
\label{sec:experiments_under_fluctuating_network}
Under fluctuating network conditions, we evaluated the effectiveness of Dynatune's dynamic tuning based on real-time network measurements.
Specifically, we varied the RTT and packet loss rate during the experiments.

\subsubsection{Adaptivity to RTT Fluctuations}
\label{sec:rtt_fluctuations}

\begin{figure*}[tb]
    \centering
    \subfloat[Gradual RTT fluctuation from 50\,ms to 200\,ms and back to 50\,ms.]{
        \includegraphics[width=0.85\linewidth]{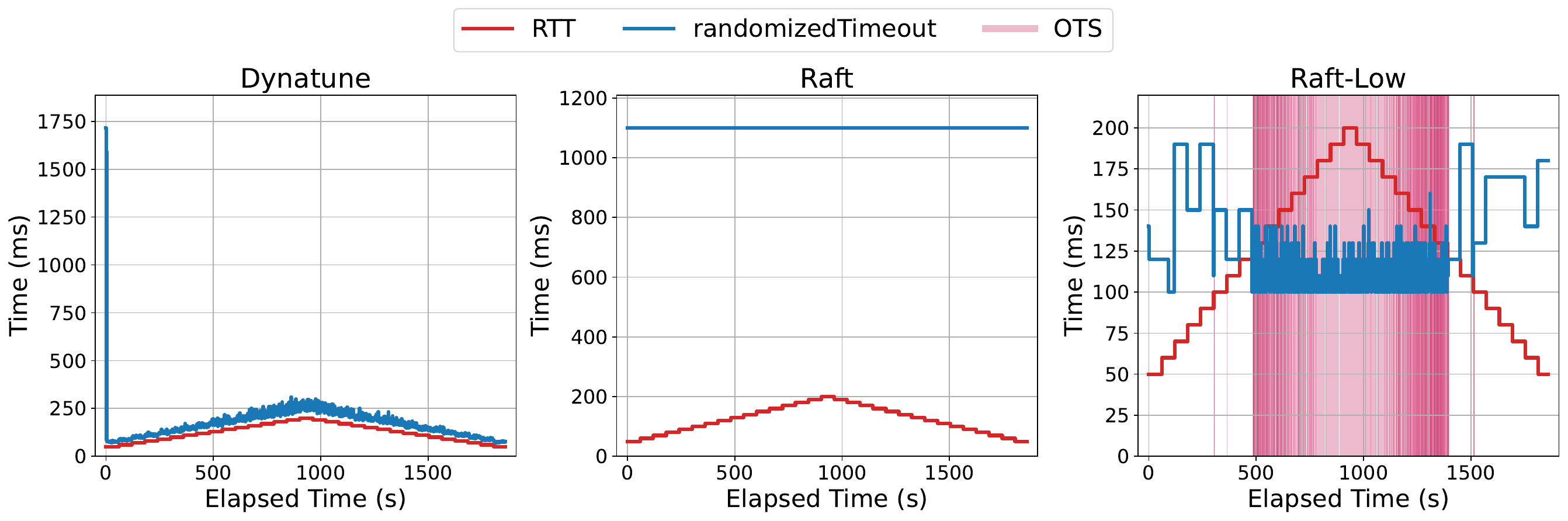}
        \label{fig:rtt_fluctuation_gradual}
    }\\
    \subfloat[Radical RTT fluctuation from 50\,ms to 500\,ms and back to 50\,ms.]{
        \includegraphics[width=0.85\linewidth]{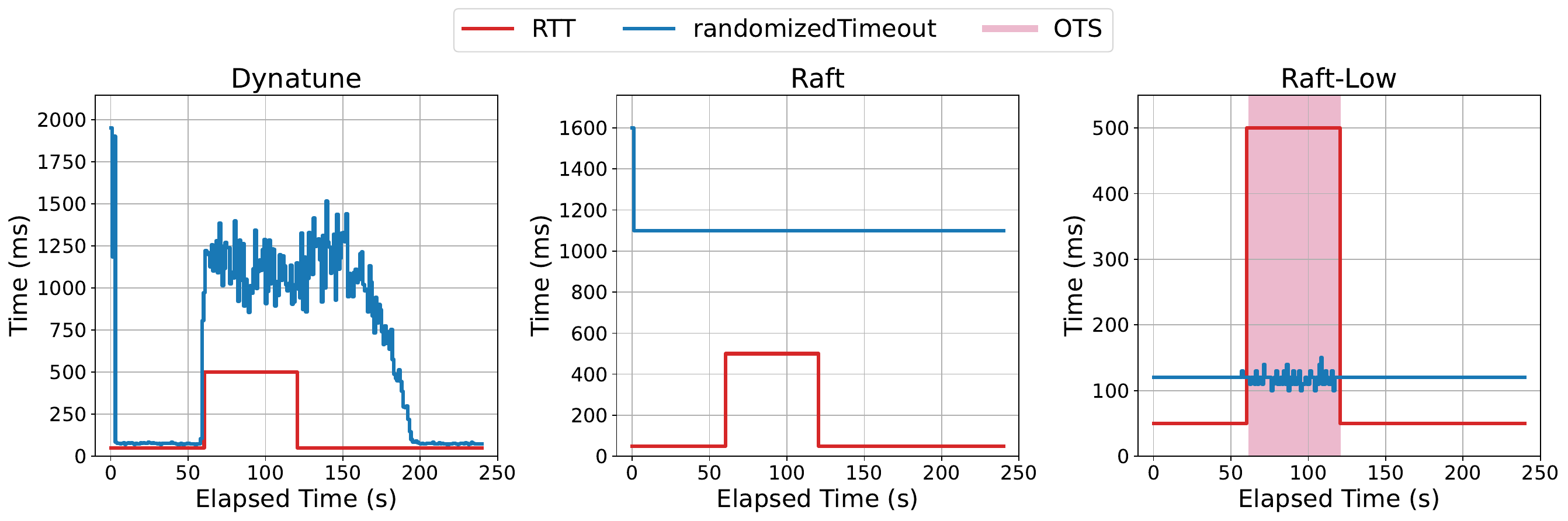}
        \label{fig:rtt_fluctuation_radical}
    }
    \caption{Comparison of the third smallest $\mathrm{randomizedTimeout}$ and OTS time for Dynatune, Raft, and Raft-Low under fluctuating RTT conditions. OTS is displayed as background shading, arising from a leader's absence caused by unnecessary elections, which occur when follower election timers time out.}
    \label{fig:rtt_fluctuation_experiment}
\end{figure*}

In this experiment, we evaluated how Dynatune adapts to both gradual and radical RTT changes.
Section~\ref{sec:election_performance} showed that Dynatune reduces the detection and OTS times by adjusting $E_t$ according to the RTT.
Here, we verified that Dynatune can maintain a small $E_t$ safely even while the RTT changes.

For the experimental setup, we compared Dynatune against Raft with the default election parameters and \emph{Raft-Low}, where we set the election parameters to 1/10 of the default values.
During the experiment, the third smallest $\mathrm{randomizedTimeout}$ value among the five servers was measured each second.
This value represents the majority threshold in this five-server setting because the pre-vote mechanism starts an election only after the election timers of a majority of the servers have expired, as described in Section~\ref{sec:implementation}.
We did not issue any requests or inject any failures, and we varied the RTT according to the two patterns.
The overall results are shown in Figure~\ref{fig:rtt_fluctuation_experiment}.

First, in the gradual fluctuation pattern, the RTT increased from 50\,ms to 200\,ms in 10\,ms steps and then decreased back to 50\,ms.
Each RTT value was maintained for one minute.
Figure~\ref{fig:rtt_fluctuation_gradual} presents the results.
Dynatune continuously adjusted $\mathrm{randomizedTimeout}$ as the RTT changed.
This behavior confirms that Dynatune controlled $E_t$ as designed, keeping it small enough to reduce the detection time.
Raft, by contrast, kept $\mathrm{randomizedTimeout}$ at approximately 1100\,ms, far above the RTT.
This conservative setting avoided false detections and unnecessary elections, but it would also delay failure detection after a leader failure and increase the OTS time.
Raft-Low, with $E_t$ fixed at 100\,ms, remained stable at first while the RTT was below 100\,ms.
As the RTT increased beyond this value, however, Raft-Low began triggering unnecessary elections, causing OTS time even though the leader had not failed.
Specifically, around 300 seconds, Raft-Low experienced a short OTS time.
It then temporarily recovered because the randomization happened to yield a larger $\mathrm{randomizedTimeout}$.
However, as the RTT continued to increase and remain above the timeout, the OTS time recurred and persisted from about 500 to 1400 seconds.

Next, in the radical fluctuation pattern, the RTT stayed at 50\,ms for 1 minute, abruptly increased to 500\,ms for 1 minute, and then returned to 50\,ms.
Figure~\ref{fig:rtt_fluctuation_radical} shows the results.
In Dynatune, the sudden increase to 500\,ms caused each follower's election timer to expire.
After the timers expired, the followers discarded the network measurement data they had gathered, became candidates, and initiated a pre-vote.
During this pre-vote, however, heartbeats from the leader arrived because the leader was still alive, causing the candidates to abort the process and revert to followers.
Thus, Dynatune experienced a false detection, but no election occurred and there was no OTS time.
When the RTT returned from 500\,ms to 50\,ms, the variance of the measured RTT increased temporarily, causing Dynatune to set a conservatively large $E_t$ during this transition.
This behavior suppressed false detections while the network was unstable, and $E_t$ decreased once the measurements stabilized.
Raft, by contrast, used a conservative timeout of approximately 1100\,ms, which maintained service availability even when the RTT increased suddenly.
However, this large timeout would delay failure detection and increase the OTS time after an actual leader failure.
For Raft-Low, $E_t$ was set to 100\,ms, making $\mathrm{randomizedTimeout}$ small during the first minute.
This configuration would enable rapid failure detection if a fault occurred.
However, from 60 to 120 seconds, the RTT increased to 500\,ms and exceeded $\mathrm{randomizedTimeout}$, causing repeated elections and preventing a leader from being elected.
During this period, the service could not process any requests owing to the absence of a leader.

\subsubsection{Adaptivity to Packet Loss Rate Fluctuations}
\label{sec:p_fluctuations}

\begin{figure*}[tb]
    \centering
    \subfloat[Heartbeat interval $h$]{
        \includegraphics[width=0.85\linewidth]{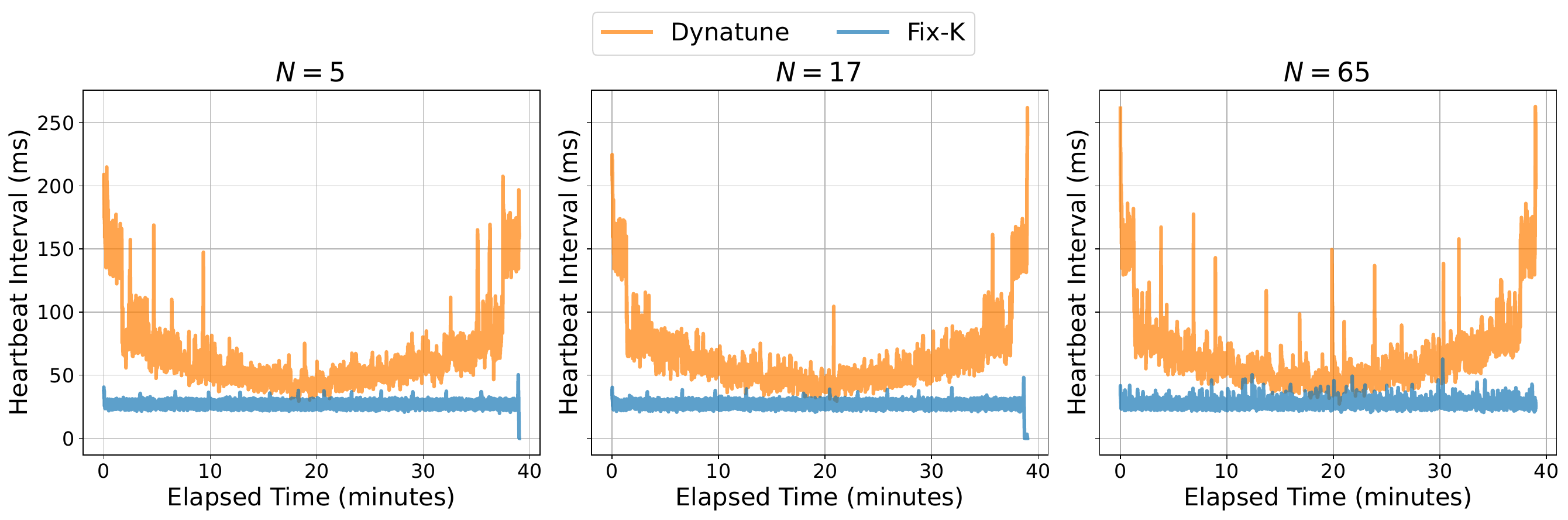}
        \label{fig:heartbeat_interval}
    }\\
    \subfloat[CPU utilization (maximum 200\% owing to 2-core allocation)]{
        \includegraphics[width=0.85\linewidth]{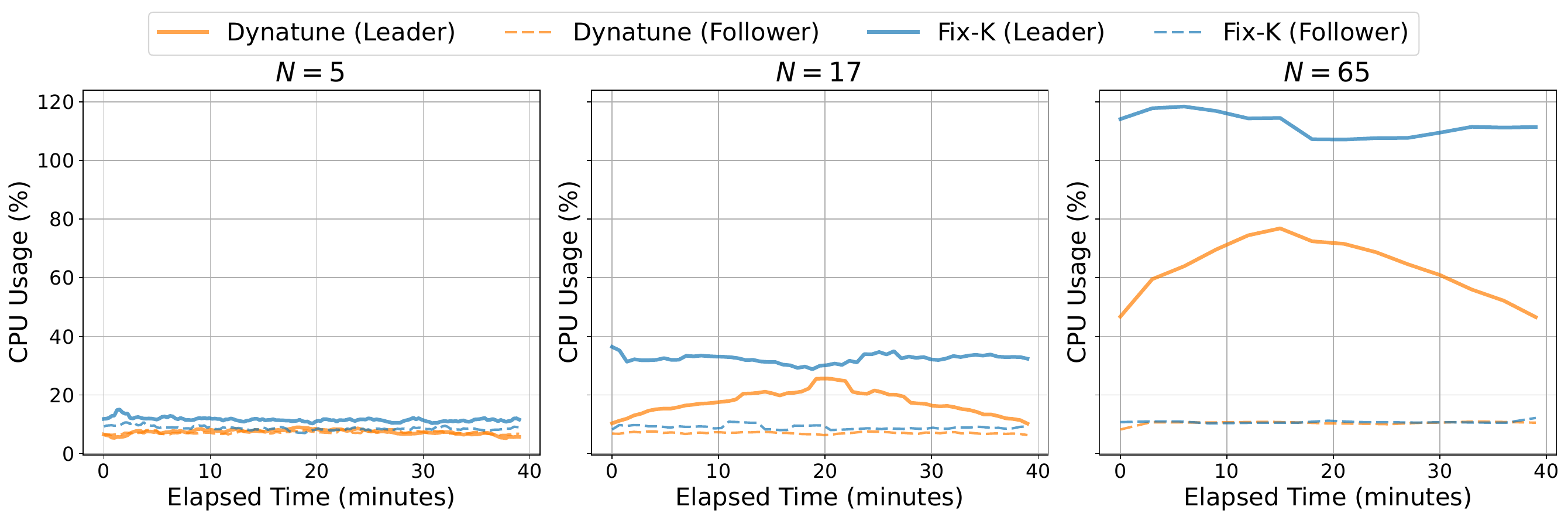}
        \label{fig:cpu_usage}
    }\\
    \subfloat[Network I/O (combined values of input and output)]{
        \includegraphics[width=0.85\linewidth]{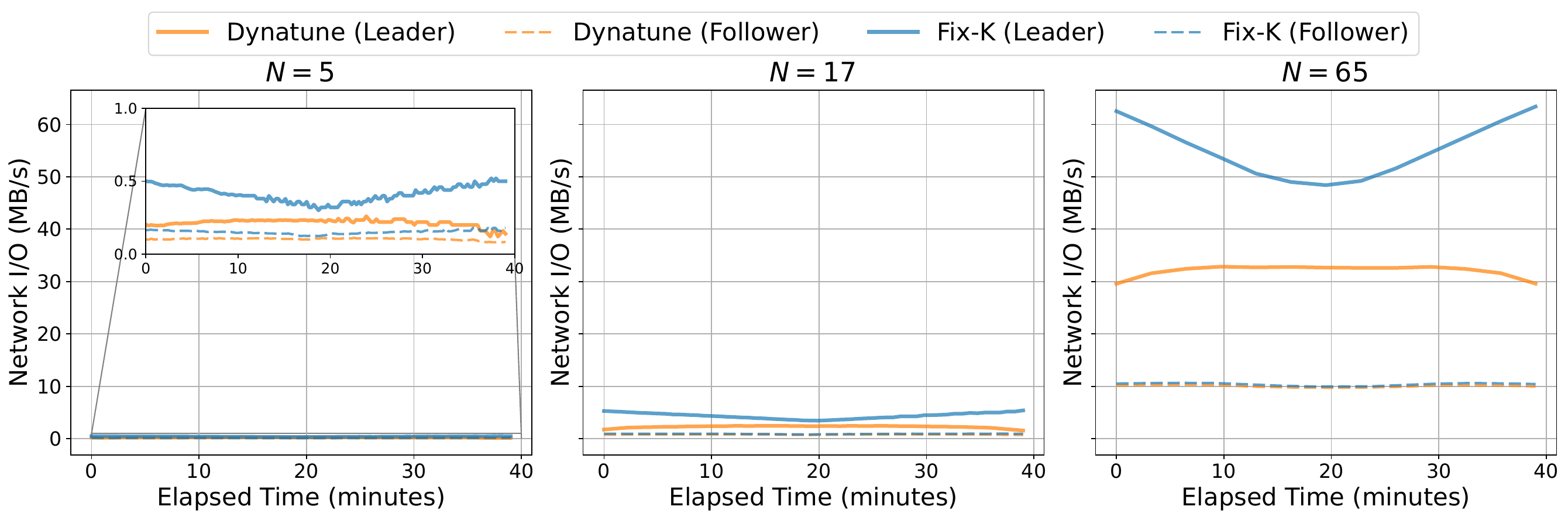}
        \label{fig:network_io}
    }
    \caption{Comparison of heartbeat interval and resource consumption between Dynatune and Fix-K under packet loss rate fluctuations for server counts of $N=5$, 17, and 65.}
    \label{fig:packet_loss_rate_fluctuation_experiment_combined}
\end{figure*}

We evaluated the effectiveness of auto-tuning the heartbeat interval $h$ in Dynatune.
We compared the standard version of Dynatune, which performed auto-tuning of $h$ to control $K$, the number of heartbeat messages sent before an election timer expires, with a version in which the tuning was disabled, referred to as \emph{Fix-K}.
In Fix-K, we always set $K$ ($=E_t/h$) to 10, based on the default settings of Raft.%

In this experiment, the RTT was fixed at 200\,ms, and the packet loss rate was increased from 0\% to 30\% and then decreased back to 0\% in increments of 5\%.
Each packet loss rate setting was maintained for 3 minutes.
We ran the experiment with the number of servers $N$ set to 5, 17, and 65, corresponding to 4, 16, and 64 followers, respectively.
Because the 65-server configuration required more machine resources than the other experiments, we used an AWS \texttt{m6a.48xlarge} instance with 192 vCPUs instead of the setup described in Section~\ref{sec:exp_setting}.
Each server was allocated two cores and 4\,GB of memory.
We recorded changes in $h$ and election events, and collected per-server CPU usage and network I/O every 5\,s from each container with \texttt{docker stats}.
Figure~\ref{fig:packet_loss_rate_fluctuation_experiment_combined} shows the overall results.

Figure~\ref{fig:heartbeat_interval} shows the changes in $h$ as the packet loss rate varied.
When the packet loss rate increased, Dynatune shortened $h$ so that heartbeats could still reach followers under high-loss conditions.
When it decreased, Dynatune lengthened $h$ to reduce resource consumption.
This behavior confirms that Dynatune adjusted $h$ according to the measured packet loss rate.

Figure~\ref{fig:cpu_usage} compares the CPU utilization of Dynatune and Fix-K, including both the leader and one follower.
The leader in Fix-K consumed more CPU resources throughout the experiment than the leader in Dynatune.
For $N=65$, the CPU utilization of the Fix-K leader frequently exceeded 100\%, showing that frequent heartbeat exchanges placed substantial processing load on the leader.
Dynatune, by contrast, used less CPU than Fix-K throughout the experiment and less than half as much CPU when the packet loss rate was low.
Its CPU utilization varied with the packet loss rate, reflecting the adjustment of heartbeat frequency.
This behavior indicates that Dynatune reduced resource use under low packet loss and sent more heartbeats under high packet loss to maintain reliable delivery.

Figure~\ref{fig:network_io} shows the aggregate network I/O during the experiment.
Fix-K consistently used more bandwidth than Dynatune.
In addition, Fix-K exhibited a valley pattern, as higher packet loss rates resulted in fewer responses from followers.
By contrast, Dynatune maintained a more consistent network I/O pattern by increasing the number of heartbeats sent as packet loss increased, ensuring that heartbeats reliably reached followers despite the higher packet loss.

Through this experiment, we observed that no unnecessary elections occurred under packet loss rates up to 30\% in either Fix-K or Dynatune.
These results confirm that Dynatune's dynamic adjustment of $h$ based on the packet loss rate effectively prevents unnecessary elections while reducing resource consumption.

\subsection{Real Distributed Experiment on AWS}
\label{sec:distributed_experiment}
\begin{figure}[tb]
        \centering
    \includegraphics[width=0.9\linewidth]{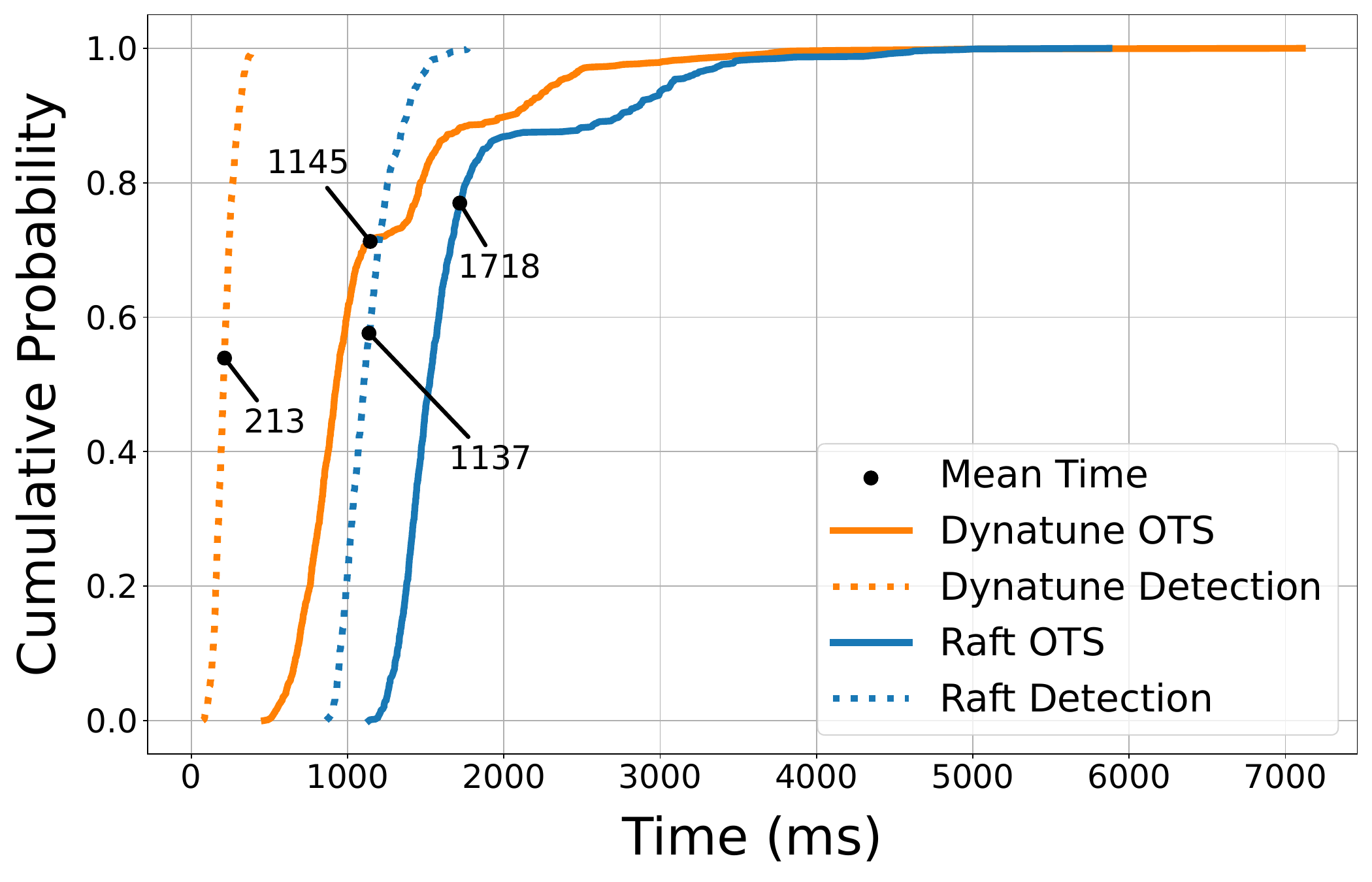}
        \caption{Cumulative probability of detection and OTS times for Dynatune vs.~Raft in the AWS distributed experiment.}
        \label{fig:distributed_experiment}
\end{figure}
We also evaluated Dynatune in a real distributed deployment to verify that it works effectively in practice.
The earlier experiments, by contrast, ran on a single physical machine to avoid timing errors in the measured detection and OTS times.
To perform this evaluation, we built a geo-replicated SMR on AWS by deploying five \texttt{m5.large} instances, each equipped with two CPU cores and 8\,GB of memory, in Tokyo, London, São Paulo, N. California, and Sydney.
On these five servers, we repeated the experiment of Section~\ref{sec:election_performance}.
Because the measurements were taken on multiple physical machines synchronized with Network Time Protocol (NTP), clock synchronization errors of several tens of milliseconds may be included.
We regard the measured times as approximate values and use them to assess whether Dynatune remains effective in such an environment.

Figure~\ref{fig:distributed_experiment} shows the results.
Focusing on the mean values, Dynatune reduced the detection time by approximately 81\% compared with Raft.
Similarly, Dynatune reduced the OTS time by approximately 33\%.
These results confirm that Dynatune detects leader failures quickly and reduces the OTS time in a real distributed environment.

\subsection{Discussion}
\label{sec:discussion}

Section~\ref{sec:election_performance} showed that Dynatune reduced the detection time by 78\% and the OTS time by 45\%.
The election time, however, was longer with Dynatune.
Raft took 249\,ms ($1420 - 1171$), whereas Dynatune took 519\,ms ($777 - 258$).
This increase was caused by the smaller $E_t$ set by Dynatune.
The resulting narrower randomization range increased the likelihood that multiple servers would start elections simultaneously.
This higher probability of split votes led to repeated election attempts.
Our approach shortens the OTS time by reducing the detection time while maintaining availability, and combining it with methods that shorten the election time~\cite{fluri_improving_2018,tang_improved_2022,fu_improved_2021,zhang_escape_2022} could address this limitation and further reduce the OTS time.

The results in Section~\ref{sec:rtt_fluctuations} show that static parameter settings make it difficult to safely shorten the OTS time.
This is particularly true in geo-replicated SMR, where server communication is unstable because of fluctuating network conditions. 
In addition, maintaining well-configured election parameters is difficult in replication methods in which servers move dynamically, as described in~\cite{kostler_fluidity_2023}. 
Therefore, a dynamic approach that adjusts election parameters according to network conditions, as in Dynatune, is necessary for effectively reducing the OTS time.

As shown in Section~\ref{sec:peak_throughput}, Dynatune and Raft achieved comparable peak throughput.
In leader-based SMR, the leader orders all client requests, so minimizing additional load on the leader is important.
In Dynatune, each follower performs the parameter computation, and the data required for network measurement are piggybacked onto heartbeat messages rather than maintained by the leader, which keeps the leader's overhead small.
In addition, as shown in Section~\ref{sec:p_fluctuations}, Dynatune adjusts the number of heartbeats according to the packet loss rate.
Since the peak throughput experiment had no packet loss, the heartbeat count remained small, and the load on the servers and the network was also minimal.
Owing to these design choices, the overhead introduced by auto-tuning is likely small under stable network conditions, allowing Dynatune to maintain throughput comparable to that of Raft.

\section{Applicability of Dynatune}
\label{sec:Applicability}

In this section, we show that Dynatune is applicable beyond Raft to leader-based consensus algorithms that employ heartbeat-based failure detection.
We take Multi-Paxos as a representative example and implement Dynatune on JPaxos~\cite{konczak_jpaxos_2011}, an open-source implementation of Multi-Paxos.
We then repeat the core experiments of Section~\ref{sec:Evaluation} on this implementation, confirming that Dynatune is also effective on JPaxos, a Multi-Paxos implementation.
Finally, we discuss its applicability to other leader-based consensus algorithms.

\subsection{Leader Failover in Multi-Paxos}
\label{sec:multipaxos_failover}

Multi-Paxos is a leader-based consensus algorithm in which a single server, called the leader, orders client requests while the other servers follow its decisions, as in Raft.
Multi-Paxos specifies only the procedure for agreeing on the order of requests.
Its safety is guaranteed regardless of the success or failure of leader election.
For liveness, a single leader is required; however, the concrete mechanisms for detecting its failure and electing a new one are not specified and are left to implementations~\cite{lamport_paxos_2001}.
Indeed, it has been noted that there is a large gap between the description of the algorithm and the requirements of a real system, and that implementers must supply many additional mechanisms~\cite{chandra_paxos_2007}.

In practice, implementations widely adopt the same approach as Raft, detecting leader failure with heartbeats and timeouts.
For example, production systems such as Google Chubby, which uses Multi-Paxos, and Ceph, which uses a Multi-Paxos-like variant, employ this approach (the mechanism is called a \emph{lease} in these systems, but the principle is the same)~\cite{burrows_chubby_2006,weil_ceph_2006}.

However, these systems are either not publicly available or are not faithful implementations of Multi-Paxos itself, and are thus not suitable as a basis for applying and evaluating our method.
We therefore target JPaxos, an open-source implementation of Multi-Paxos.
In the following, we describe JPaxos's failure detection and election mechanisms, contrasting them with Raft.

The failure detection mechanism is almost identical to that of Raft.
In JPaxos, the leader sends an \texttt{Alive} message to all other servers at a fixed interval $\mathrm{FDSendTimeout}$, which corresponds to the heartbeat interval $h$ in Raft.
If a server does not receive the next \texttt{Alive} within $\mathrm{FDSuspectTimeout}$ after the last one, it regards the leader as failed and proceeds to elect a new leader.
This $\mathrm{FDSuspectTimeout}$ corresponds to Raft's election timeout $E_t$.
Accordingly, in JPaxos, Dynatune dynamically tunes $\mathrm{FDSuspectTimeout}$ and $\mathrm{FDSendTimeout}$.

The leader election mechanism differs from that of Raft.
In Raft, a follower whose $E_t$ expires increments the term and becomes a candidate, and the server that collects a majority of votes becomes the new leader.
If multiple followers become candidates at nearly the same time, the votes may split and no candidate obtains a majority.
In that case, the election is retried.
To avoid such split votes, Raft randomizes $E_t$ so that candidates are less likely to appear simultaneously.

JPaxos adopts a view-based leader election scheme derived from Viewstamped Replication~\cite{oki_viewstamped_1988,liskov_viewstamped_2012}.
In JPaxos, each of the $n$ servers is assigned a unique identifier from $0$ to $n-1$, and elections are managed with a monotonically increasing number called a view.
Each view has a unique leader.
The leader of view $v$ is deterministically fixed as the server whose identifier equals $v \bmod n$.
Suppose that the current view is $v$ and the server with identifier $i$ suspects the leader.
This server proposes the smallest view $v' > v$ satisfying $v' \bmod n = i$, that is, the next view in which it becomes the leader.
It sends a \texttt{Prepare} message for view $v'$ and becomes the new leader if it collects \texttt{PrepareOK} messages from a majority of servers.

View numbers are totally ordered, and each server supports only the largest view it has seen so far.
Consequently, even if multiple servers suspect the leader simultaneously, the server proposing the largest view is uniquely selected, and a split vote as in Raft does not arise.
Randomizing $E_t$ to avoid split votes is therefore unnecessary in JPaxos.

\subsection{Implementation on JPaxos}
\label{sec:jpaxos_implementation}

We forked JPaxos\footnote{\url{https://github.com/JPaxos/JPaxos}} and integrated Dynatune, and the source code is publicly available online\footref{footnote:dynatune_url}.

The network measurement and parameter tuning are implemented as described in Section~\ref{sec:Dynatune}.
As noted above, $\mathrm{FDSuspectTimeout}$ and $\mathrm{FDSendTimeout}$ correspond to $E_t$ and $h$, respectively.
We add the fields $\langle \mathrm{ID}, \mathrm{ts}, \mathrm{RTT}, h \rangle$ to the \texttt{Alive} and \texttt{AliveReply} messages.
The former corresponds to the heartbeat, and the latter does not exist in the original JPaxos but was introduced because Dynatune requires followers to reply to each \texttt{Alive} message.
JPaxos sends messages smaller than a certain size over UDP.
Since the sizes of \texttt{Alive} and \texttt{AliveReply} messages are under this threshold, these messages are exchanged over UDP.

There are two differences from the Raft-based implementation.
The first is the randomization of the election timeout, which exists in Raft but not in JPaxos, as described above.
The second is the pre-vote phase, which exists in etcd but not in JPaxos.
Apart from these, the core of Dynatune (RTT estimation, computation of $E_t$ and $h$, and follower-driven tuning) is unchanged from the Raft version.
We also provide the same four runtime arguments introduced in the Raft version (the safety factor $s$, the probability $x$, $\mathrm{minListSize}$, and $\mathrm{maxListSize}$).

\subsection{Experiments}
\label{sec:jpaxos_experiments}

We repeat the three core experiments of Section~\ref{sec:Evaluation} with Dynatune implemented on JPaxos.
These experiments evaluate failover performance, peak throughput, and adaptivity to RTT fluctuations.
The hardware and network environment is identical to that of Section~\ref{sec:exp_setting}, with etcd replaced by JPaxos.
We compare a statically configured JPaxos baseline with JPaxos integrating Dynatune, referring to the former as \emph{JPaxos} and the latter as \emph{Dynatune}.
We do not quantitatively compare the results with those of the Raft version, because the aim of these experiments is to confirm applicability, and because the base implementations differ.
To align the experimental conditions with the Raft experiments, we set $E_t$ ($\mathrm{FDSuspectTimeout}$) $= 1000$\,ms and $h$ ($\mathrm{FDSendTimeout}$) $= 100$\,ms for both configurations.
We set the runtime arguments of Dynatune to $s = 2$, $x = 0.999$, $\mathrm{minListSize} = 10$, and $\mathrm{maxListSize} = 100$, and used the JPaxos default values for all remaining parameters.

\subsubsection{Failover Performance}
\label{sec:jpaxos_failover}

We intentionally failed the leader 1000 times and measured the detection time and OTS time for JPaxos and Dynatune.
The RTT between servers was fixed at 100\,ms, the packet loss rate at 0\%, and jitter was not added.

\begin{figure}[tb]
    \centering
    \includegraphics[width=0.9\linewidth]{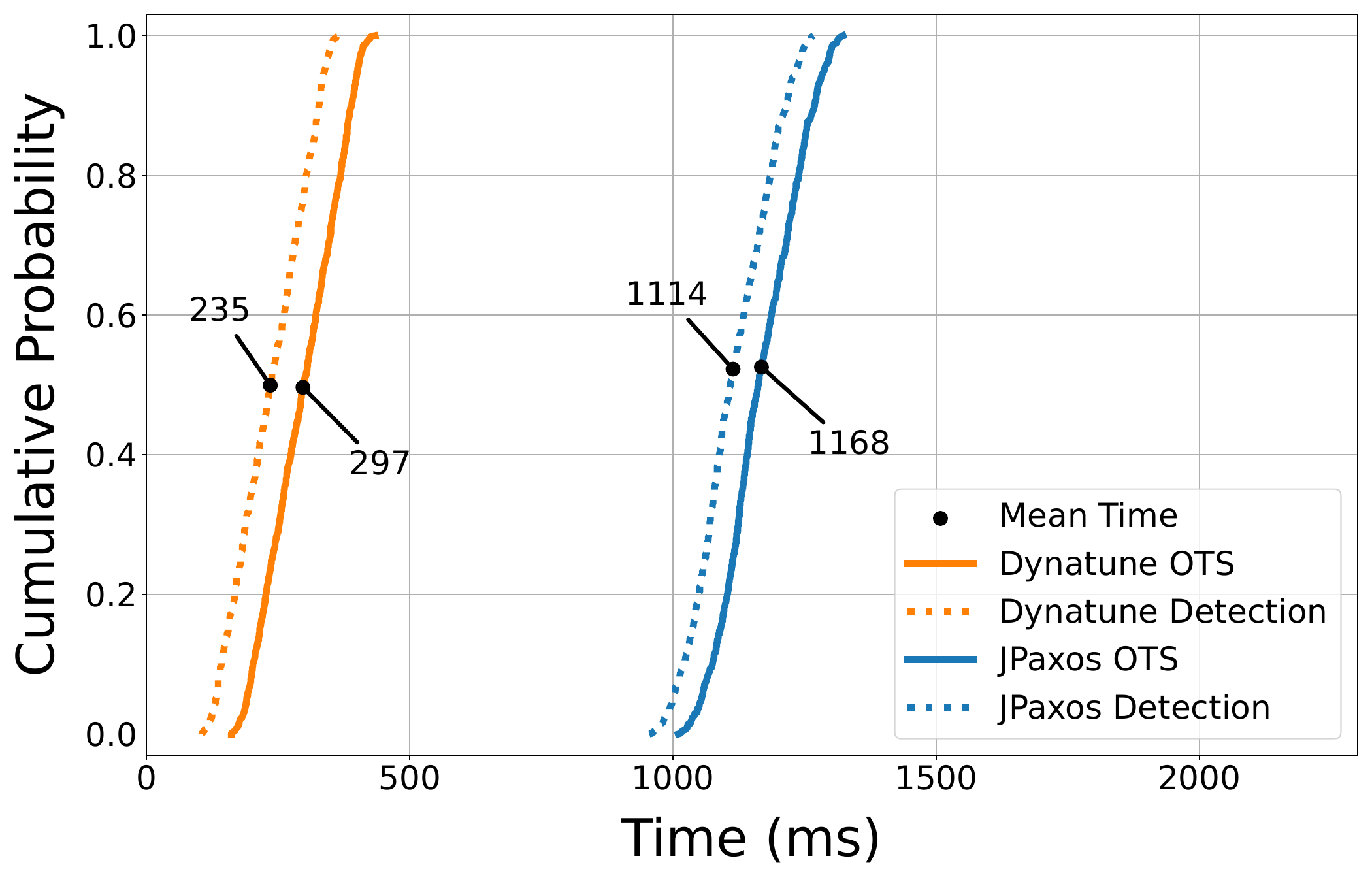}
    \caption{Cumulative probability of detection and OTS times for Dynatune vs.~JPaxos.}
    \label{fig:jpaxos_election_performance}
\end{figure}

Figure~\ref{fig:jpaxos_election_performance} shows the experimental results.
Dynatune reduced the detection time by 79\%, from 1114\,ms to 235\,ms, and the OTS time by 75\%, from 1168\,ms to 297\,ms, compared with the JPaxos baseline.
As in the Raft version, the detection time was shortened owing to the small $E_t$ tuned by Dynatune.
Moreover, because split votes do not occur in JPaxos's view-based election, the substantial increase in election time observed for Raft in Section~\ref{sec:discussion} did not occur.

\subsubsection{Peak Throughput}
\label{sec:jpaxos_throughput}

We measured peak throughput using the procedure described in Section~\ref{sec:peak_throughput}.
We used increments of 1000\,req/s up to 6000\,req/s and finer increments of 300\,req/s from 6000 to 7500\,req/s because latency rose steeply in this region.
This experiment assesses the overhead that Dynatune's automatic tuning adds to request processing.

\begin{figure}[tb]
    \centering
    \includegraphics[width=0.8\linewidth]{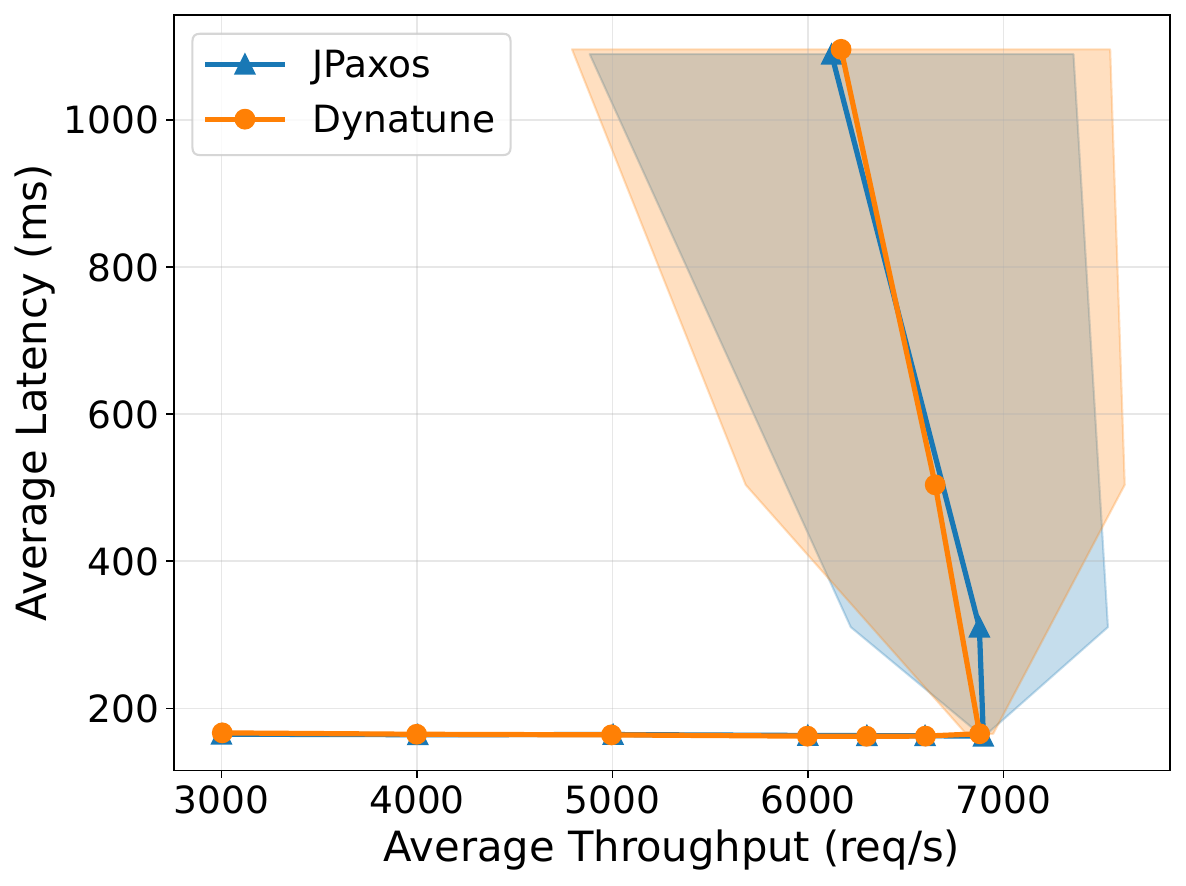}
    \caption{Throughput and latency for Dynatune vs.~JPaxos.}
    \label{fig:jpaxos_peak_throughput}
\end{figure}

Figure~\ref{fig:jpaxos_peak_throughput} shows the average latency, average throughput, and standard deviation of the throughput.
When comparing peak throughput, JPaxos achieved 6897\,req/s, whereas Dynatune achieved 6878\,req/s, indicating that their peak throughputs are comparable.

\subsubsection{Adaptivity to RTT Fluctuations}
\label{sec:jpaxos_rtt}

We evaluated adaptivity to RTT fluctuations under gradual and radical fluctuation patterns.
The gradual pattern increased the RTT from 50\,ms to 200\,ms and then decreased it to 50\,ms, whereas the radical pattern abruptly increased the RTT from 50\,ms to 500\,ms and then returned it to 50\,ms.
We evaluated only Dynatune because the JPaxos baseline uses a fixed $\mathrm{FDSuspectTimeout}$ of 1000\,ms and therefore does not adapt to RTT changes.
During the experiment, we measured the time series of the $\mathrm{FDSuspectTimeout}$ values tuned by the five servers and the occurrence of unnecessary elections.

\begin{figure*}[tb]
    \centering
    \includegraphics[width=0.8\linewidth]{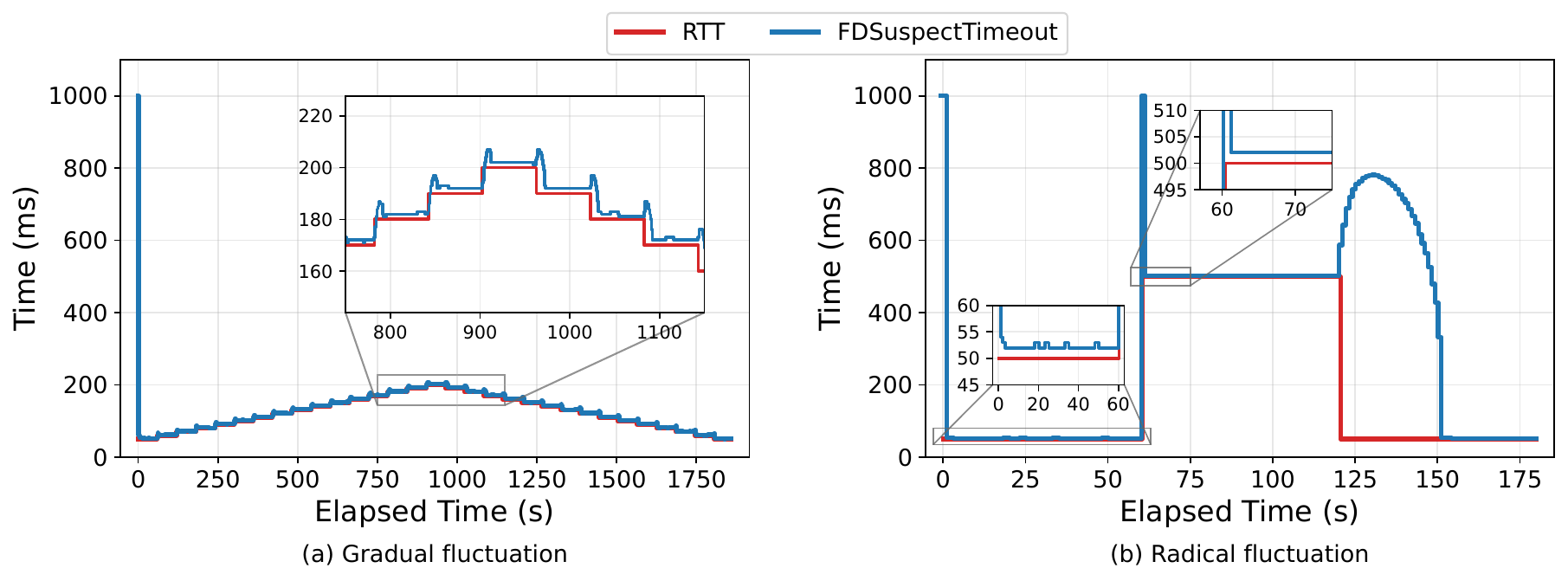}
    \caption{Time series of RTT and $\mathrm{FDSuspectTimeout}$ for Dynatune on JPaxos under gradual and radical RTT fluctuation patterns.}
    \label{fig:jpaxos_rtt_fluctuation}
\end{figure*}

In the gradual fluctuation pattern (Figure~\ref{fig:jpaxos_rtt_fluctuation}a), because JPaxos does not randomize $\mathrm{FDSuspectTimeout}$, the timeout value remained close to the RTT.
As the RTT increased in 10\,ms steps, $\mathrm{FDSuspectTimeout}$ momentarily fell below the current RTT.
However, the next heartbeat arrived before the timeout fired and updated $\mathrm{FDSuspectTimeout}$ to a larger value, so no election occurred.

In the radical fluctuation pattern (Figure~\ref{fig:jpaxos_rtt_fluctuation}b), during the first 60\,s when the RTT was 50\,ms, Dynatune tuned $\mathrm{FDSuspectTimeout}$ to approximately 52\,ms.
When the RTT then abruptly increased to 500\,ms, no heartbeat could arrive within this 52\,ms window, and the timeout fired, triggering a false leader-failure detection.
Unlike the Raft version, JPaxos does not implement a pre-vote phase, so the election was not suppressed and one election occurred.
The cost of this unnecessary election was one RTT of communication, during which the cluster had no leader.
After that, $\mathrm{FDSuspectTimeout}$ adapted to track the 500\,ms RTT, and when the RTT returned to 50\,ms, $\mathrm{FDSuspectTimeout}$ followed accordingly.

Figure~\ref{fig:jpaxos_rtt_fluctuation}b and the Dynatune result in Figure~\ref{fig:rtt_fluctuation_radical} show that the re-convergence time after the RTT returned from 500\,ms to 50\,ms was shorter in JPaxos than in the Raft-based implementation.
Immediately after the RTT decrease, the RTT sample window temporarily contained samples from both the 500\,ms and 50\,ms periods, increasing the measured standard deviation and causing Dynatune to set $\mathrm{FDSuspectTimeout}$ to a conservative value.
As samples from the 50\,ms period accumulated, samples from the 500\,ms period were replaced, and $\mathrm{FDSuspectTimeout}$ returned to a smaller value.
In JPaxos, $\mathrm{FDSuspectTimeout}$ re-converged in roughly 30\,s, whereas the Raft-based implementation took roughly 70\,s under the same RTT pattern.
This difference arises because timeout randomization in Raft affects the calculation of $h$.
In Raft, the actual election timer is randomized between $E_t$ and $2E_t$, so its expected value is 1.5 times $E_t$.
Because $h$ is calculated from this timer value, the Raft-based implementation sets a larger $h$, which slows the update of the RTT sample window.
By contrast, JPaxos does not randomize the timeout, so $h$ is smaller and the RTT sample window is updated more quickly.

\subsection{Discussion}
\label{sec:jpaxos_discussion}

Across the three experiments, Dynatune behaved on JPaxos as it did on the Raft-based implementation.
It shortened the detection and OTS times, kept the peak throughput comparable to the baseline, and tracked the RTT changes, with only a single unnecessary election under the abrupt tenfold increase.
Dynatune therefore applies to Multi-Paxos as well as Raft and reduces the OTS time.

Dynatune is a general approach that applies independently of implementation details.
Raft and JPaxos differ in several concrete ways, such as the presence of a pre-vote phase, the leader-election mechanism, and whether the election parameters are randomized.
Dynatune nonetheless applied to both and reduced the detection and OTS times in each.
The reason is that Dynatune does not modify these details and tunes only the heartbeat interval and the election timeout, which are common to leader-based algorithms in which the leader sends periodic heartbeats and a follower starts a new election after a timeout.

Beyond Raft and Multi-Paxos, Dynatune is also broadly applicable to other leader-based SMR systems, such as Chubby and Viewstamped Replication.
Chubby~\cite{burrows_chubby_2006}, a representative implementation of Multi-Paxos, designates one server as the master and manages mastership through a lease mechanism.
When the lease expires, another server initiates an election to select a new master.
The master periodically submits a dummy ``heartbeat'' value~\cite{chandra_paxos_2007} to refresh the lease and indicate its liveness.
This heartbeat renewal plays the same role as heartbeat messages in Raft, while the renewal frequency and lease timeout correspond to Raft's heartbeat interval and election timeout, respectively.
Similarly, Viewstamped Replication~\cite{oki_viewstamped_1988, liskov_viewstamped_2012} also employs periodic leader messages and timeout-driven view changes, and applying Dynatune could improve the responsiveness of failure detection and leader reconfiguration.

\section{Related Work}
\label{sec:Related_Work}

SMR replicates service state across multiple servers and uses consensus algorithms to maintain consistency by ensuring that the servers agree on the same order of client requests, thereby providing fault tolerance.
Among these algorithms, leader-based approaches such as Raft and Multi-Paxos are widely used in practice for their simple and clearly structured design.
Because of these properties, they have been adopted in many services, including etcd~\cite{etcd_webpage}, TiDB~\cite{huang_tidb_2020}, CockroachDB~\cite{taft_cockroachdb_2020}, Chubby~\cite{burrows_chubby_2006}, and Ceph~\cite{weil_ceph_2006}.

Leader-based consensus algorithms have a common limitation that causes services to temporarily stop processing client requests when the leader fails.
The OTS time is a critical factor affecting service availability.

Several studies aim to reduce the OTS time in leader-based consensus algorithms.
For Raft, Fluri et al.~\cite{fluri_improving_2018} studied how packet loss and network partitioning affect the election time, and introduced a timeout policy that lengthens the timeout after every split vote, which shortens the election time.
Tang et al.~\cite{tang_improved_2022} divided the servers into ordinary and prioritized groups and directed votes toward the prioritized ones, reducing split votes and thereby shortening the election time.
Fu et al.~\cite{fu_improved_2021} and Zhang et al.~\cite{zhang_escape_2022} modified Raft's voting mechanism so that an election completes in a single round, which also reduces the election time.
All of these efforts aim to reduce the election time, whereas we reduce the detection time by tuning the election parameters according to the measured network conditions.

Many studies have also examined election parameters in leader-based consensus algorithms.
For example, Iosif et al.~\cite{iosif_leadership_2020} randomized the election parameters so that every server has an equal probability of becoming the leader.
Choumas et al.~\cite{choumas_using_2022} tuned the parameters to make a specific server more likely to be elected, which reduces request latency during normal operation.
Huang et al.~\cite{huang_performance_2020} modeled the probability that the network partitions under high packet loss rates and used the model to derive a suitable ratio between the election timeout and the heartbeat interval.
These studies mainly address responsiveness during normal operation, whereas our goal is to reduce the OTS time.

Wang~\cite{wang_ballast_2025} proposed BALLAST, which shares our motivation of reducing OTS time by adaptively controlling election timeout.
While Dynatune directly tunes parameters using explicit measurements of RTT and packet loss rate, BALLAST selects timeout values from a predefined candidate set based on observed outcomes such as election success and recovery behavior.
In that study, Wang compared BALLAST with static and heuristic timeout policies, and Dynatune-inspired methods derived from the conference version of this work~\cite{shiozaki_dynatune_2025}.
Specifically, \texttt{dynatune\_et} adapts the election timeout while keeping heartbeat intervals fixed, whereas \texttt{dynatune\_joint} adapts both the election timeout and heartbeat interval.
The results showed that BALLAST outperformed static and heuristic timeout policies, while the Dynatune-inspired methods achieved the best performance among the compared approaches: \texttt{dynatune\_joint} yielded the shortest \textit{recovery\_time}, and \texttt{dynatune\_et} yielded the lowest \textit{unwritable\_fraction}.
These findings suggest that the effectiveness of the Dynatune-style approach has also been observed in subsequent related work.

\section{Conclusion}
\label{sec:Conclusion}
In this paper, we proposed Dynatune, which dynamically tunes the election parameters according to network conditions.
Dynatune measures the RTT and packet loss rate through heartbeats and adjusts the election parameters based on these measurements.
Dynatune can therefore detect leader failures in a timely manner and safely shorten the OTS time even under fluctuating network conditions.
We implemented Dynatune on Raft and Multi-Paxos, and then evaluated it against statically configured baselines under various fault scenarios and network conditions.
Across the two implementations, Dynatune reduced the detection time by 78--79\% and the OTS time by 45--75\%, while maintaining availability even under fluctuating network conditions.

As future work, we will investigate integrating Dynatune with existing approaches for reducing election time to further shorten the OTS time.
We will also explore the applicability of Dynatune to Byzantine fault-tolerant algorithms such as PBFT~\cite{castro2001practical}.
At present, Dynatune is vulnerable to malicious servers that may report falsified metrics, potentially triggering unnecessary elections and causing a loss of liveness.
To address this issue, we will focus on developing mechanisms to sanitize and validate network metrics.

\section*{Acknowledgment} 
The authors used ChatGPT (GPT-5.2) provided by OpenAI to assist with English language editing. 
All technical content was produced by the authors, and all AI-suggested revisions were carefully reviewed to ensure accuracy and consistency with the intended meaning.

\bibliographystyle{IEEEtran}
\bibliography{references}

\begin{IEEEbiography}[{\includegraphics[width=1in,height=1.25in,clip,keepaspectratio]{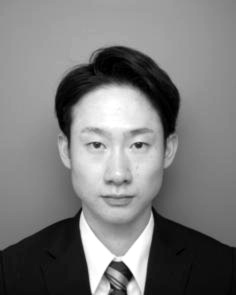}}]{Kohya Shiozaki}
Kohya Shiozaki received the B.E. and M.E. degrees from Toyohashi University of Technology, Aichi, Japan, in 2023 and 2025, respectively. He is currently a Ph.D. student at Toyohashi University of Technology. His research interests include distributed algorithms and distributed systems.
\end{IEEEbiography}

\begin{IEEEbiography}[{\includegraphics[width=1in,height=1.25in,clip,keepaspectratio]{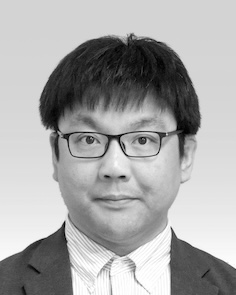}}]{Junya Nakamura} (M'19) received the B.E.~and M.E.~degrees from Toyohashi University of Technology, Aichi, Japan, in 2006 and 2008, respectively, and the Ph.D.~degree in Information Science and Technology from Osaka University, Osaka, Japan, in 2014.
He is currently an associate professor of Information and Media Center, Toyohashi University of Technology.
His research interests include theoretical and practical aspects of distributed algorithms and systems.
He is a member of ACM, IEEE, IEEE Computer Society, IEICE, and IPSJ.
\end{IEEEbiography}

\EOD
\end{document}